\documentclass[prb,reprint,superscriptaddress,twocolumn]{revtex4-2}

\usepackage[dvipdfmx]{graphicx}
\usepackage{bm}
\usepackage{amsmath,amssymb,bm,braket,mathrsfs,mathtools}
\usepackage{color}

\DeclareMathOperator{\tr}{tr}

\DeclareMathOperator{\diag}{diag}

\begin{document}
\title{Majorana multipole response: General theory and application to wallpaper groups}
\author {Shingo Kobayashi}
\affiliation{RIKEN Center for Emergent Matter Science, Wako, Saitama 351-0198, Japan}
\author{Yuki Yamazaki}
\affiliation{Department of Physics, Nagoya University, Nagoya 464-8602, Japan}
\author{Ai Yamakage}
\affiliation{Department of Physics, Nagoya University, Nagoya 464-8602, Japan}
\author{Masatoshi Sato}
\affiliation{Yukawa Institute for Theoretical Physics, Kyoto University, Kyoto 606-8502, Japan}

\date{\today}
\begin{abstract} 
Whereas identification of Cooper pair symmetry is the first and crucial step in the investigation of unconventional superconductors, only a few have been established so far because of its own difficulties. 
To solve this problem, we develop a theory for identification of pairing symmetry using knowledge of topological superconductivity. 
 Establishing the multipole theory of emergent Majorana fermions in time-reversal-invariant topological superconductors, we discover a one-to-one correspondence between the electromagnetic response of Majorana fermions and Cooper pair symmetry.
The emergent Majorana fermions host magnetic structures that share the same irreducible representation with Cooper pairs under crystalline symmetry.  
We furthermore reveal that Majorana fermions in high-spin or nonsymmorphic superconductors may exhibit magnetic octupole responses, which give a direct evidence of these exotic superconducting states.
Electric responses of multiple Majorana Kramers pairs are also clarified.
Our theory provides the fundamentals for identification of unconventional Cooper pairings through surface-spin-sensitive measurements as well as that for manipulation of Majorana fermions by external electromagnetic fields.

\end{abstract}
\maketitle

\section{Introduction}
Over the past decade, tremendous progress has been made in understanding of topological phases of matter. 
Comprehensive classifications based on 
the $K$-theory~\cite{Schnyder08,Kitaev09,Schnyder09,Ryu10,Morimoto13,Chiu13,Shiozaki14,Shiozaki15,Shiozaki16,Chiu16,Shiozaki17,CFang17,Shiozaki18atiyah,Cornfeld19,Shiozaki19classification,Okuma19,Song19topological} and the symmetry-indicator~\cite{Bradlyn17topological,Kruthoff17,Po17,ZSong18quantitative,Khalaf18,Elcoro2020magnetic} have been pushed forward to search for new topological phases enabled by crystalline symmetry.
In particular, by incorporating the first-principle calculations and material databases, the symmetry-indicator method has uncovered several thousands of topologically nontrivial materials~\cite{Zhang19catalogue,Vergniory19complete,Tang19sciadv,Tang2019efficient,DWang19,Xu2020high}.   

While these classifications can be extended to topological superconductors (TSCs)~\cite{Ono19,Skurativska20,Shiozaki19variants,Geier20,Ono2020refined, Ono2020mathbbz2enriched}, the search for TSCs faces its own difficulties which do not exist in other topological materials:  
Whereas the classifications rely on  Cooper pair symmetry,  the identification of the latter is very difficult. In fact, in spite of a lot of effort, the exact Cooper pair symmetry has yet to be determined in many unconventional superconductors, with a few exceptions such as high-$T_{\rm c}$ cuprates. 
Because the typical energy scale of unconventional superconducting gaps is much smaller than that of insulating gaps, the experimental means to identify the pairing symmetry is limited.
For instance, the angle-resolved photoemission spectroscopy, which is commonly used to identify topological materials, is not available, because its resolution has not reached the energy scale of the superconducting gap. 
Moreover, there is no established theory for the  prediction of Cooper pair symmetry. 
The first-principle calculation, which is powerful for the prediction of topological insulators, 
has not been reliable yet for the calculation of unconventional pairing states.
Therefore, a new principle to identify the pairing symmetry is highly desired.

In this paper, we show that Cooper pair symmetry in time-reversal invariant TSCs is directly measured by electromagnetic responses of Majorana fermions (MFs) on their surfaces.
The emergent MFs appear as surface zero energy Andreev bound states \cite{Hu94,Kashiwaya2000,Volovik03, Sato09, Wilczek09, 
Hasan10,Qi11,Tanaka12,Alicea12,Ando2015,SatoFujimoto16,Mizushima16,Sato17,yonezawa2016bulk} and have been paid the most attention as a potential candidate for fault-tolerant qubits for topological quantum computation~\cite{Nayak08}. 
The increased interest in the emergent MFs offers proposals of versatile time-reversal invariant TSCs, such as superconducting doped topological insulators~\cite{Hor10,Fu10,Sasaki11,Sasaki12,Hashimoto15,Fu14,Matano16,Yonezawa17}  and Dirac semimetals~\cite{Aggarwal16,Wang16,Kobayashi15PRL,Hashimoto16,Oudah16,Kawakami18,Zhang19,Kawakam2019}. 
The MFs in time-reversal invariant TSCs commonly form Kramers pairs at zero energy, which we dub Majorana Kramers pairs (MKPs). They host the spin degrees of freedom ensured by time-reversal symmetry (TRS), which constitutes electromagnetic structures unique to the emergent MFs. 
The existence of electromagnetic structures provides possibility of electromagnetic responses even though the emergent MFs are electrically neutral. 
In addition, TSCs often host topological invariants protected by crystalline symmetry ~\cite{Chiu13,Morimoto13,Shiozaki14,Benalcazar14,Shiozaki16}, and MKPs receive an additional constraint from them. The electromagnetic structures associated with MKPs turn out to acquire versatile structures and realize an anisotropic magnetic response~\cite{SatoFujimoto09,Chung09,Nagato09,Shindou10,Mizushima12,Tsutsumi13,Mizushima16,Xiong17,Volpez19,Kobayashi19,Yamazaki19,Plekhanov2021,Yamazaki20}. Such magnetic anisotropy is a salient feature of the emergent MFs, since neither electric nor magnetic responses are possible for elementary Majorana particles~\cite{Kayser83,Radescu85,Boudjema89}.

To prove the relation between electromagnetic structures and Cooper pair symmetry, we establish a general theory of electromagnetic structures of MKPs. 
Then, using the general theory, we exhaustively classify the electromagnetic structures under crystalline symmetry. We consider all possible minimal set of MFs located at any of the highest symmetry points on surface Brillouin zones (BZs). 
Our theory depends only on the irreducible representations (irreps) of MKPs and Cooper pairs, and it can calculate responses of MKPs protected by $\mathbb{Z}$ and $\mathbb{Z}_2$ invariants~\cite{Kobayashi19,Yamazaki19,Yamazaki20} in a unified way. The results show that the one-to-one correspondence between irreps of magnetic couplings and those of Cooper pairs always holds whenever only a single MKP exists. 
The irreps of magnetic structures are manifest in possible electromagnetic multipole responses.
The emergent MFs show magnetic dipole order, but they also exhibit magnetic octupole one in special situations.  
There are two types of mechanisms for the magnetic octupole: (i) spin 3/2 symmetry induced magnetic octupole and (ii) nonsymmorphic symmetry induced magnetic octupole. 
Whereas the former one was partially discussed in our previous study~\cite{Kobayashi19},  we here generalize it to all surface MFs with wallpaper groups. 
We also discover the latter mechanism of the magnetic ocutupole order, which is realized by  MKPs protected by glide symmetry. 
These results indicate that the magnetic octupole responses provide a direct evidence of the exotic superconductivity in high-spin or non-symmorphic superconductors.

The above one-to-one correspondence does not hold when multiple MKPs exist on a surface.
Even in this case, however, we can determine the paring symmetry through the responses of MFs.
In addition to the magnetic responses, the multiplicity of MFs enables the electric responses. 
One may specify Cooper pair symmetry by combining these two responses.

This paper is organized as follows. In Sec.~\ref{sec:general}, we develop a general theory of electromagnetic responses of MFs in time-reversal invariant TSCs.
First, we summarize relevant symmetries in Sec.~\ref{sec:pre}. 
Then, in Sec.~\ref{sec:wigner}, we introduce the topological invariants for surface MFs by combining the group theoretical method with the $K$-theroy classification. 
In Sec.\ref{sec:degeneracy}, we count the minimal degeneracy of surface MFs required by crystalline symmetry.
In most cases, the minimal degeneracy is the Kramers one imposed by time-reversal symmetry (TRS), but the four-fold degeneracy is required at the $\bar{M}$ point for $pgg$ and $p4g$.
We evaluate quantum operators of MFs and determine the leading electromagnetic couplings of MFs in Secs.~\ref{sec:multipole_op} and \ref{sec:emcoupling}.
In particular, we prove that a nonzero quantum operator of a single MKP must be a magnetic operator and shares the same irrep with the gap function. (See Eq.(\ref{eq:corr}).)
In Secs.~\ref{sec:application} and \ref{sec:multiple}, we apply the general theory to MFs protected by the wallpaper groups. 
Our results for a single MKP are summarized in Tables \ref{tab:EAZsym} and \ref{tab:EAZnonsym}.
We also discuss magnetic ocutupole responses in high-spin and nonsymmorphic superconductors in Secs.~\ref{sec:high} and~\ref{sec:nonsym}, respectively.   
In Sec~\ref{sec:model}, we illustrate the magnetic octupole responses by using concrete models. 
In Sec.~\ref{sec:multiple}, we clarify electromagnetic structures of double MKPs realized at the $\bar{M}$ point of $pgg$ and $p4g$, where 
the additional multiplicity enables electric structures.
We first point out that MFs belong to short representations for particle-hole symmetry (PHS) or chiral symmetry (CS), because of the self-antiparticle nature of MFs.
Then, by constructing the short representation explicitly, we evaluate the electromagnetic responses of the double MKPs, which are summarized in Table \ref{tab:elerep}. 
In Sec.~\ref{sec:summary}, we provide a summary and 
 discuss the experimental relevance.

\section{Summary of results}
\label{sec:summary_of_results}
Our main technical accomplishment is a multipole theory of MKPs, which allows us to determine electromagnetic structure of MKPs from only crystalline symmetry and Cooper pair symmetry in bulk superconductors. As elaborated in Sec.~\ref{sec:multipole_op}, the formulae for determining the electromagnetic structures of MKPs are given by Eqs.~(\ref{eq:chafull}) and (\ref{eq:chiOmega+-}): $\chi^{\Omega}_g$ is the character of representations on MKPs for a given group $g \in G_0$ and $\chi^{\Omega^{\pm}}_g$ is the character of representations decomposed into electric ($+$) and magnetic ($-$) structures. 
Before going into the technical details, we summarize main results of the multipole theory and connections to physical systems. 

{\it Electromagnetic structures of MKPs.}---Applying the formulae~(\ref{eq:chafull}) and (\ref{eq:chiOmega+-}) to the wallpaper groups which are space group symmetry preserved on a surface and focusing on a minimal MKP lying at a high-symmetry point on the surface BZ, the electromagnetic structures of MKPs are exhaustively classified in Table~\ref{tab:EAZsym} and~\ref{tab:EAZnonsym}. The key findings are summarized as follows.
\begin{itemize}
\item For a single MKP, only a magnetic structure is allowed, and it shares the same irrep with the Cooper pair under crystalline symmetry, i.e., $\chi^{\Omega}_g =\chi^{\Omega^{-}}_g =\eta_g$ [Eq.~(\ref{eq:corr})], with $\eta_g$ in Eq.~(\ref{eq:eta_g}).
\item A large majority of a single MKP have a magnetic dipole; namely, the response function is given by a linear function of a magnetic field, $f(\bm{B}) \sim \bm{B} \cdot \bm{n}$ with $\bm{n}$ being specified by crystalline symmetry (see Table~\ref{tab:EAZsym} and~\ref{tab:EAZnonsym}). Systems showing the magnetic dipole response include $^3$He-B phase~\cite{Chung09,Nagato09,Shindou10,Mizushima12,Tsutsumi13,Mizushima16}, superconducting doped topological insulators/semimetals~\cite{Ando2015,Sato17,yonezawa2016bulk}, and nodal superconductors with crystalline symmetry-protected Majorana flat bands~\cite{SAYang14,Mizushima14odd,Kobayashi14,Kobayashi18,Hu2018majorana}. 
\item A single MKP exhibits a magnetic octupole response if the surface symmetry is $p6$, $p3m1$, $p31m$, or $p6m$ and its spin is $3/2$; see Fig.~\ref{fig:p6model}.  The response function is of the order of $\mathcal{O}(|\bm{B}|^3)$ since the linear terms are prohibited by the crystalline symmetry. Material candidates realizing the magnetic octupole response are high-spin superconductors such as half-Hausler compounds~\cite{Goll08,Butch11,Tafti13,GXu16,Bay12,Brydon16,Kim18} and antiperovskite Dirac metals~\cite{Oudah16,Kawakami18}; topological superconducting states realize MFs with spin 3/2. The magnetic octupole response in the half-Hausler compound YPtBi has been demonstrated in Ref.~\onlinecite{Kobayashi19}.  
\item A single MKP shows another magnetic octupole response if the surface symmetry is $pmg$ or $pgg$ (nonsymmorphic) and the MKP emerges at a high symmetry point on the surface BZ boundary; see Fig.~\ref{fig:pmmamodel}. 
The magnetic response is again of the order of $\mathcal{O}(|\bm{B}|^3)$, but the expression of the response function is different from the previous one; see Secs.~\ref{sec:high} and~\ref{sec:nonsym} for more discussions. The candidates are glide symmetry-protected topological superconductors such as UCoGe~\cite{Daido2019,Yoshida2019}.
\item  A single MKP is forbidden and double MKPs are realized if the surface symmetry is $pgg$ or $p4g$ and the MF appears at the $\bar{M}$ point on the surface BZ. The response function of the double MKPs consists of a mixture of several irreps, including electric response; see Sec.~\ref{sec:multiple} and Fig.~\ref{tab:elerep} for more details. Our multipole theory is applicable to multiple MFs and enables us to distinguish electric structures from magnetic ones in a systematic way.
\end{itemize}

{\it Connection to physical observables.}---The electromagnetic structure of MKPs represents internal degrees of freedom of MFs such as spin and orbital and thus can be measured through the coupling to external fields such as magnetic fields, strains, and so on; see Sec.~\ref{sec:emcoupling} for more discussions.
 In particular, a single MKP hosts only a magnetic structure, whose irrep one-to-one corresponds to that of Cooper pairs. Thus, we can determine bulk pairing symmetries from surface magnetic responses; the magnetic structures can be probed via surface-spin-sensitive measurements such as spin-resolved tunneling spectroscopy~\cite{Jeon17,Cornils17}, spin relaxation rate~\cite{Chung09}, spin susceptibility~\cite{Nagato09}, thermal conductivity~\cite{Nakai14,Xie15,Gnezdilov16} under magnetic fields, and so on. For instance, the spin susceptibility is enhanced in a particular direction due to the anisotropic spin structures of the MKP~\cite{Nagato09,Mizushima14odd}. The anisotropy is directly linked to  Cooper pair symmetry under crystalline symmetry.

\section{General theory}
\label{sec:general}

\subsection{Symmetries}
\label{sec:pre}
Three-dimensional (3d) time-reversal invariant TSCs host helical Majorana fermions on their surfaces, which are ensured by the so-called 3d winding number~\cite{Grinevich88, Schnyder08, Sato09, Sato10}. While the 3d winding number is defined only for fully-gapped TSC, its parity can be defined even for nodal superconductors~\cite{Sasaki11} and ensures the existence of an odd number of MKPs. 
In this paper, we focus our attention on how MKPs respond to external electromagnetic fields. Obviously, electric fields only extract moderate responses from a MKP as they maintain TRS. On the other hand, magnetic fields substantially affect MKPs because the 3d winding number and its parity are ill-defined under TRS breaking external fields. However, this does not imply that MKPs are unstable under any magnetic field. In real systems,
MKPs are also protected by crystalline symmetry when they are located at a high symmetry point or line in the surface BZ. In fact, crystalline symmetry provides an additional topological invariant that stabilizes MKPs. Therefore, even if the 3d winding number and its parity are ill-defined, MKPs cannot respond to magnetic fields so much as long as the crystalline symmetry for the additional topological invariant is maintained. Namely, only magnetic fields that break the crystalline symmetry may destabilize MKPs.

First, we summarize symmetries considered in this paper. 
We consider space groups that are compatible with surfaces hosting MKPs.
The corresponding space groups are wallpaper groups, which consist of 17 groups: $p1$, $p2$, $pm$, $pg$, $cm$, $pmm$, $pmg$, $pgg$, $cmm$, $p4$, $p4m$, $p4g$, $p3$, $p3m1$, $p31m$, $p6$, and $p6m$. 
In addition, we take into account TRS $T$, PHS $C$, and their combination, CS $\Gamma=-iTC$. These symmetries form a group $G$, which is decomposed into 
\begin{align}
G=G_0+TG_0+CG_0+\Gamma G_0,
\end{align}
where $G_0$ is a wallpaper group. 
The group $G$ acts on the Bogoliubov-de Genne (BdG) Hamiltonian 
\begin{align}
H({\bm k})=
\begin{pmatrix}
h({\bm k}) & \Delta({\bm k})\\
\Delta^{\dagger}({\bm k}) & -h^T(-{\bm k})		
\end{pmatrix},
\end{align}
where $h({\bm k})$ and $\Delta({\bm k})$ are a normal Hamiltonian and a gap function, respectively.
For $g\in G_0$, the action for the BdG Hamiltonian  reads
\begin{align}
{\cal U}_g^{\bm k} H({\bm k}) {\cal U}_g^{{\bm k} \dagger} =H(g{\bm k}),
\quad
{\cal U}_g^{\bm k}=
\begin{pmatrix}
U_g^{\bm k} & \\
& \eta_g U_g^{-{\bm k}*}
\end{pmatrix},
\end{align}
where $U_g^{\bm k}$ is  a unitary operator obeying
\begin{align}
U_g^{\bm k} h({\bm k})U_g^{{\bm k} \dagger}=h(g{\bm k}), 
\quad U_g^{\bm k}\Delta({\bm k})U_g^{-{\bm k}T}=\eta_g \Delta(g{\bm k}), \label{eq:eta_g}
\end{align}
with a $U(1)$ factor $\eta_g$ determined by the pairing symmetry of the gap function. 
($\eta_g$ must be $\pm 1$ when the gap function has TRS.)
For TRS $T$ and PHS $C$, we have
\begin{align}
&{\cal U}_T H^*({\bm k}) {\cal U}_T^{\dagger}=H(-{\bm k}),
\quad 
{\cal U}_T=
\begin{pmatrix}
U_T & \\
& U_T^*
\end{pmatrix},
\nonumber\\
&{\cal U}_C H^*({\bm k}){\cal U}_C^{\dagger}=-H(-{\bm k}),
\quad
{\cal U}_C=
\begin{pmatrix}
 &1 \\
1& 
\end{pmatrix},
\label{eq:TC}
\end{align}
where $U_T$ is a unitary operator for TRS on the normal Hamiltonian with 
$
U_T U_T^*=-1.
$
Using the complex conjugation operator $K$, we can also write Eq.(\ref{eq:TC}) as
\begin{align}
{\cal T} H({\bm k}) {\cal T}^{-1}=H(-{\bm k}), \quad
{\cal C} H({\bm k}) {\cal C}^{-1}=-H(-{\bm k}),
\end{align}
with ${\cal T}={\cal U}_T K$ and ${\cal C}={\cal U}_C K$.

The unitary operator ${\cal U}_g^{\bm k}$ for $g\in G$ provides a projective representation of $G$,
\begin{align}
z_{g,h}^{gh{\bm k}}{\cal U}_{gk}^{\bm k}=\left\{
\begin{array}{ll}
{\cal U}_g^{h{\bm k}}{\cal U}_h^{\bm k} & \mbox{if $g$ is unitary}\\
{\cal U}_g^{h{\bm k}}({\cal U}_h^{{\bm k}})^*& \mbox{if $g$ is anti-unitary}
\end{array}
\right.,
\end{align}
where $z_{g,h}^{\bm k}$ is a $U(1)$ phase called factor system.
The factor system is given as follows:
Let $g=\{p|{\bm a}_p\}$ be an element of $G_0$, where $p$ is a point group operation and ${\bm a}_p$ is a translation: $\{p|{\bm a}_p\}, {\bm x} \mapsto p{\bm x}+{\bm a}_p$.
The product of $g=\{p|{\bm a}_p\}$ and $g'=\{p'|{\bm a}_{p'}\}$ reads
\begin{align}
\{p|{\bm a}_p\}\{p'|{\bm a}_{p'}\}&=\{pp'|p{\bm a}_{p'}+{\bm a}_p\}
\nonumber\\
&=\{e|p{\bm a}_{p'}+{\bm a}_p-{\bm a}_{pp'}\}\{pp'|{\bm a}_{pp'}\},
\end{align}
where $\varepsilon$ is the identity operator. 
Correspondingly, the factor system for $g,g'\in G_0$ is given by
\begin{align}
z_{g,g'}^{\bm k}=z_{p,p'}e^{-i{\bm k}\cdot(p{\bm a}_{p'}+{\bm a}_p-{\bm a}_{pp'})},
\end{align}
where $z_{p,p'}=\pm 1$ originates from the double projective representation of spin rotation in $p$ and $p'$, and the exponential factor is the Bloch factor of $\{e|p{\bm a}_{p'}+{\bm a}_p-{\bm a}_{pp'}\}$.
We also require that the subscript $g$ of ${\cal U}_g^{\bm k}$ obeys the linear representation of $G$ where $T$ and $C$ commute with any $g\in G$ and obey $T^2=C^2=\{e|{\bm 0}\}$. Here $\{e|{\bm 0}\}$ is the unit element. 
By combining this property with commutation relations in $G$, we can determine the factor system for other elements in $G$ uniquely. 
For instance, for $g=\{p|{\bm a}_p\}\in G_0$, we have
\begin{align}
{\cal U}_g^{-{\bm k}}{\cal U}_T={\cal U}_T({\cal U}_g^{\bm k})^*, 
\quad
{\cal U}_g^{-{\bm k}}{\cal U}_C=\eta_g{\cal U}_C({\cal U}_g^{\bm k})^*, 
\label{eq:gC}
\end{align}
which lead to
\begin{align}
z^{\bm k}_{g,T}=z_{T,g}^{\bm k}, \quad 
z^{\bm k}_{g, C}=z_{C, g}^{\bm k},
\end{align}
because of ${\cal U}_{Tg}^{\bm k}={\cal U}_{gT}^{\bm k}$ and ${\cal U}_{Cg}^{\bm k}={\cal U}_{gC}^{\bm k}$.
Furthermore, from $T^2=-1$ and $C^2=1$, we have 
\begin{align}
z^{\bm k}_{T,T}=-1, \quad z^{\bm k}_{C,C}=1,
\end{align}
since ${\cal U}_{T^2}={\cal U}_{C^2}={\cal U}_{\{e|{\bm 0}\}}$.

\subsection{Wigner's test and 1d topological invariants}
\label{sec:wigner}

When crystalline symmetry is taken into account, the topological classification  is diversely ramified, and crystalline symmetry-protected topological phases appear~\cite{Chiu13,Morimoto13,Shiozaki14,Benalcazar14,Shiozaki16}. Crystalline symmetry-protected topological invariants defined in low-dimensional subspaces may protect MKPs. 
In particular,  MKPs at high symmetry points on surface BZs are supported by crystalline symmetry-protected 1d topological invariants.

To see this, let us consider a high symmetry point ${\bm k}$ of $G$ on a surface BZ, which is the projection of a high symmetry line $l_{\bm k}$ in the bulk BZ, and the little group $G^{\bm k}$ that keeps the high symmetry point ${\bm k}$ (and the high symmetry line $l_{\bm k}$) invariant. 
The little group $G^{\bm k}$ is decomposed into
\begin{align}
G^{\bm k}=G_0^{\bm k}+TG_0^{\bm k}+CG_0^{\bm k}+\Gamma G_0^{\bm k},
\end{align}
where $G_0^{\bm k}$ consists of all elements in $G_0$ belonging to $G^{\bm k}$.
On the 1d subspace $l_{\bm k}$, any element in $G_0^{\bm k}$ commutes with the BdG Hamiltonian $H({\bm k})$. 
Thus, if we take the basis where ${\cal U}_g^{\bm k}$ ($g\in G_0^{\bm k}$) is decomposed into irreps of $G_0^{\bm k}$, 
\begin{align}
{\cal U}_g^{\bm k}=\oplus_{\alpha} 
\begin{pmatrix}
U_g^{{\alpha}} & \\
& \eta_g (U_g^{\alpha})^*
\end{pmatrix} \label{eq:decompU}
\end{align}
where $\alpha$ labels the irreps, then $H({\bm k})$ on $l_{\bm k}$ is also decomposed into subsectors,
\begin{align}
H({\bm k})=\oplus_{\alpha}H^{\alpha}({\bm k}), \label{eq:decompBdG}
\end{align}
where $H^{\alpha}({\bm k})$ is a Hamiltonian belonging to the irrep $\alpha$.
The set of 1d Hamiltonians $H^{\alpha}({\bm k})$ on $l_{\bm k}$ defines  crystalline symmetry-protected 1d topological invariants.

To identify the 1d topological invariant of $H^{\alpha}({\bm k})$, we employ the Wigner test~\cite{Bradley72,Wigner59,Herring37,Inui90,Shiozaki18,Sumita19}.
The Wigner test specifies an Altland-Zirnbauer (AZ) symmetry class of $H^{\alpha}({\bm k})$, which we call  emergent AZ (EAZ) class, and determines a possible 1d topological invariant of $H^{\alpha}({\bm k})$.
For the Wigner test of $H^{\alpha}({\bm k})$, we calculate three indices $(W^T, W^C, W^{\Gamma})$ defined as follows,
\begin{align}
 &W_{\alpha}^{T} \equiv \frac{1}{|G^{\bm k}_0|} \sum_{g \in G^{\bm k}_0} 
 z^{\bm k}_{Tg,Tg} \tr [U^{\alpha}_{(Tg)^2}]=\pm 1, 0,
 \label{eq:WT} \\
 &W_{\alpha}^{C} \equiv \frac{1}{|G^{\bm k}_0|} \sum_{g \in G^{\bm k}_0} z^{\bm k}_{Cg,Cg} \tr [U^{\alpha}_{(Cg)^2}]=\pm 1, 0,
 \label{eq:WC} \\
& W_{\alpha}^{\Gamma} \equiv \frac{1}{|G^{\bm k}_0|} \sum_{g \in G^{\bm k}_0} \frac{z^{\bm k}_{\Gamma,\Gamma^{-1}g\Gamma}}{z^{\bm k}_{g,\Gamma}} \tr[U^{\alpha}_{\Gamma^{-1} g \Gamma}]^{\ast} \tr[U^{\alpha}_g]
=1,0,
\label{eq:WG} \end{align}
where $|G^{\bm k}_0|$ represents the number of elements in $G^{\bm k}_0$, and $U^{\alpha{\bm k}}_g$ is the irrep $\alpha$ of ${\cal U}_g^{\bm k}$. 
The indices $(W_{\alpha}^{T},W_{\alpha}^{C},W_{\alpha}^{\Gamma})$ indicates the presence and/or absence of TRS, PHS, and CS in $H^{\alpha}({\bm k})$ and identify the EAZ class, as shown in Table~\ref{tab:EAZ}.
Then, regarding $H^{\alpha}({\bm k})$ on $l_{\bm k}$ as a 1d system in the EAZ class, 
we can specify the possible 1d topological invariant. See Appendix~\ref{app:top_inv} also.

\begin{table}[tb]
\caption{
EAZ classes $(W_{\alpha}^{T},W_{\alpha}^{C},W_{\alpha}^{\Gamma})$ and associated 1d topological invariants
}
\label{tab:EAZ}
\begin{tabular}{ccccc}
\hline\hline
$W_{\alpha}^{T}$ & $W_{\alpha}^{C}$& $W_{\alpha}^{\Gamma}$ &EAZ class & $1$ dim.\\
\hline 
$0$ & $0$ & $0$ & A & $0$\\
$0$ & $0$ & $1$ & AIII &  $\mathbb{Z}$\\
$1$ & $0$ & $0$ & AI &  $0$\\
$1$ & $1$ & $1$ & BDI &  $\mathbb{Z}$ \\
$0$ & $1$ & $0$ & D &  $\mathbb{Z}_2$\\
$-1$ & $1$ & $1$ & DIII & $\mathbb{Z}_2$\\
$-1$ & $0$ & $0$ & AII &  $0$\\
$-1$ & $-1$ & $1$ & CII & $2 \mathbb{Z}$\\
$0$ & $-1$ & $0$ & C &  $0$\\
$1$ & $-1$ & $1$ & CI &  $0$\\
\hline\hline
\end{tabular} 
\end{table}  

For instance, consider one of the highest symmetry points of $p2$.
At the highest symmetry point,  $G_0^{\bm k}$ is $p2$ itself, {\it i.e.} $G^{\bm k}_0=p2=\{\{e|\bm{0}\},\{2_z| \bm{0}\}\}$, where $2_z$ is  two fold rotation around the $z$-axis.
The irreps of ${\cal U}_g^{\bm k}$ ($g\in G_0^{\bm k}$) are the double-valued representations $^i \bar{E}$ $(i=1,2)$, each of which corresponds to spin up and down states, respectively (we adopt the notation of irreps in the Bilbao Crystallographic Server~\cite{Elcoro17}).
TRS satisfies $[\{2_z| \bm{0}\},T]=0$ and $T^2=-\{e|\bm{0}\}$, so it follows from Eq.~(\ref{eq:WT}) that
\begin{align}
 W_{^i\bar{E}}^{T} &=  \frac{1}{2} \left( -\tr [U^{^{i}\bar{E}}_{\{e|\bm{0}\}^2}]+\tr [U^{^i\bar{E}}_{\{2_z| \bm{0}\}^2}] \right) \notag \\
 &=\frac{1}{2} \left(-1+1\right)=0.
\end{align}
Similarly, we can calculate  Eq.~(\ref{eq:WC}).  It is necessary to pay attention to the commutation relation between $C$ and $\{2_z| \bm{0}\}$, which depends on the irrep of the gap function. 
The gap function is a single-valued representation of $p2$, which is either A ($\eta_{\{2_z|{\bm 0}\}}=1$) or B ($\eta_{\{2_z|{\bm 0}\}}=-1$). For the A gap function, Eq.~(\ref{eq:gC}) implies $[C,\{2_z|\bm{0}\}]=0$, while for the B gap function,  it implies $\{C,\{2_z|\bm{0}\}\}=0$. Thus, Eq.~(\ref{eq:WC}) becomes
\begin{align}
 W_{^i\bar{E}}^{C} &=  \frac{1}{2} \left( \tr [U^{^i\bar{E}}_{\{e|\bm{0}\}^2}]-\eta_{\{2_z|{\bm 0}\}} 
 \tr[U^{^i\bar{E}}_{\{2_z| \bm{0}\}^2}] 
 \right) \notag \\
 &=\frac{1}{2} \left(1-\eta_{\{2_z|{\bm 0}\}} \right)=\begin{cases} 0 & \text{for A gap function} \\ 1 &\text{for B gap function}\end{cases}.
\end{align}
From $\Gamma =-iCT$, Eq.~(\ref{eq:WG}) is similarly obtained as
\begin{align}
  W_{^i\bar{E}}^{\Gamma} &=  \frac{1}{2} \left(|\tr [U^{^i\bar{E}}_{\{e|\bm{0}\}}]|^2+\eta_{\{2_z|{\bm 0}\}}  |\tr[U^{^i\bar{E}}_{\{2_z| \bm{0}\}}]|^2\right) \notag \\
  &=\frac{1}{2} \left(1+\eta_{\{2_z|{\bm 0}\}} \right)=\begin{cases} 1 & \text{for A gap function}\\ 0 &\text{for B gap function}\end{cases}.
\end{align}
Therefore, the EAZ class is AIII for the A gap function and D for the B gap function, respectively. 

These results are understood as follows.  Because ${^1\bar{E}}$
 and ${^2\bar{E}}$ are the $+i$ and $-i$ eigensectors of ${\cal U}_{\{2_z|{\bm 0}\}}$, respectively,   their bases are the eigenstates $|+\rangle$ and $|-\rangle$  defined by ${\cal U}_{\{2_z|{\bm 0}\}}|\pm \rangle=\pm i|\pm \rangle$.
For the A gap function,  both ${\cal T}$ and ${\cal C}$ commutes with ${\cal U}_{\{2_z|{\bm 0}\}}$ at the highest symmetry point. Therefore,  we have
\begin{align}
{\cal U}_{\{2_z|{\bm 0}\}}{\cal T}|\pm\rangle=\mp i {\cal T}|\pm\rangle,
\quad
{\cal U}_{\{2_z|{\bm 0}\}}{\cal C}|\pm\rangle=\mp i {\cal C}|\pm\rangle,
\end{align}
which implies that TRS and PHS exchange the eigensectors.
Hence, neither $H^{^1\bar{E}}({\bm k})$ nor $H^{^2\bar{E}}({\bm k})$ keeps these symmetries, and  they only retain the combination of ${\cal T}$ and ${\cal C}$, namely CS. 
Thus they belong to class AIII, as was shown by the Wigner test above.
A similar argument works for the B gap function. In this case, 
${\cal C}$ anticommutes with ${\cal U}_{\{2_z|{\bm 0}\}}$ at the highest symmetry point,  while ${\cal T}$ commutes with ${\cal U}_{\{2_z|{\bm 0}\}}$.
As a result, $H^{^1\bar{E}}({\bm k})$ and $H^{^2\bar{E}}({\bm k})$  have PHS,  and thus they belong to class D.

\subsection{Degeneracy}
\label{sec:degeneracy}
Because of TRS with $T^2=-1$, Majorana fermions at high symmetry points form Kramers pairs.
For instance, in the above $p2$ case, $^1\bar{E}$ and $^2\bar{E}$ are related to each other by TRS,  and thus once $H^{^1\bar{E}}({\bm k})$ becomes topologically non-trivial, so is $H^{^2\bar{E}}({\bm k})$. 
Therefore, the resulting Majorana modes appear in a pair.
In general, for a given irrep $\alpha$ at a high symmetry point, 
there are three possible realizations of a MKP;
(i) a MKP formed with a different irrep $\bar{\alpha}$,  (ii) a MKP formed within $\alpha$, 
and (iii) a MKP formed with another $\alpha$.
In the case (i), the EAZ class of $\alpha$ does not have TRS,  while in the latter two cases (ii) and (iii),  it hosts TRS.
Moreover,  in the case (ii), TRS of the EAZ class is bosonic, while in the case (iii), TRS of the EAZ class is fermionic. 
For instance, the EAZ class at the $\bar{X}$ point of $pg$ with $A$ gap function is  DIII. 
(See Table \ref{tab:EAZsym}.)
Thus a MKP at the $\bar{X} \equiv (\pi,0)$ point is formed between two identical irreps $(X_i, X_i)$ ($i=1,2$) (see Table~\ref{tab:nonsym}).  

The above result implies that if the irrep $\alpha$ in cases (i) and (iii) is two-dimensional, 
MFs at the high symmetry point exhibit four-fold degeneracy.
As we will see later, such additional degeneracy occurs at the $\bar{M} \equiv (\pi,\pi)$ point of $pgg$ and $p4g$.

\subsection{Majorana operators}
\label{sec:multipole_op}
To study electromagnetic structures of MFs, 
we consider a general local quantum operator $\hat{\cal O}$ defined by
\begin{align}
 \hat{\mathcal{O}} ({\bm x}) &\equiv \sum_{ij} \hat{c}_i^{\dagger}(\bm{x}) O_{ij} \hat{c}_j (\bm{x}) \notag \\
                             &= \frac{1}{2} \hat{\Psi}^{\dagger} (\bm{x})\mathcal{O} \hat{\Psi} (\bm{x})
                             \notag \\
  &= \frac{1}{2} \hat{\Psi}^{T} (\bm{x}){\cal U}_C\mathcal{O} \hat{\Psi} (\bm{x})                           
                             \label{eq:q-op}
\end{align}
with $\hat{\Psi}^T_i(\bm{x}) = (\hat{c}_i(\bm{x}),\hat{c}_i^{\dagger}(\bm{x}))$ and
\begin{align}
 \mathcal{O} = \begin{pmatrix} O & 0 \\ 0 & -O^t \end{pmatrix},
\end{align}
where the hermiticity of $O$ implies $\{\mathcal{C}, \mathcal{O}\}=0$. 
For instance, for $O=s_i$ with the Pauli matrices $s_i$, Eq.~(\ref{eq:q-op}) represents a magnetic dipole momentum of electrons.
Below, we show that nonzero $\hat{\cal O}$s for MKPs are subject to crystalline symmetry, and they determine electromagnetic responses of MKPs.

We first perform the mode expansion to extract the contribution to $\hat{\cal O}$ from MKPs at a high symmetry point ${\bm k}$ on a surface BZ.
We decompose the quantum field $\hat{\Psi}({\bm x})$ into MFs $| u_0^{(a)}({\bm x}) \rangle$ at ${\bm k}$ and others:  
\begin{align}
\hat{\Psi}({\bm x}) = \sum_a \hat{\gamma}_a | u_0^{(a)}({\bm x}) \rangle + \cdots, \label{eq:modeexp}
\end{align}
where $a$ labels MFs, and $\hat{\gamma}_a$ are Majorana operators. 
Here $\hat{\gamma}_a$ satisfies
\begin{align}
 \hat{\gamma}_a^{\dagger} &= \sum_{b} \hat{\gamma}_b (C_{\gamma})_{ba}^{\ast}, 
 \quad \{\gamma_a,\gamma_b\}=(C_{\gamma})_{ab},  
 \label{eq:gamma-rel}
\end{align}
with $(C_{\gamma})_{ba} \equiv \langle u_0^{(b)} | \mathcal{C} u_0^{(a)} \rangle$. 
(see Appendix~\ref{app:gamma}). 
Note that Eq.~(\ref{eq:gamma-rel}) reduces to the well-known Majorana condition 
$\gamma^{\dagger}_a = \gamma_a$ 
if we impose the additional constraint $ | \mathcal{C} u_0^{(a)} \rangle =  | u_0^{(a)} \rangle$. 
 Substituting Eq.~(\ref{eq:modeexp}) into Eq.~(\ref{eq:q-op}), we have
 \begin{align}
 \hat{\mathcal{O}}_{\rm MF}  ({\bm x})
 =\frac{1}{2} \sum_{a,b} \hat{\gamma}_a \hat{\gamma}_b \; \tr \left[  \mathcal{O} | u_{0}^{(b)} ({\bm x}) \rangle \langle \mathcal{C} u_{0}^{(a)} ({\bm x}) | \right],\label{eq:OMF-full}
 \end{align}
 then we separate symmetric and antisymmetric parts of Majorana operators in Eq.~(\ref{eq:OMF-full}). Since $\{\gamma_a,\gamma_b\}=(C_{\gamma})_{ab}$ is a constant, the symmetric part does not give a coupling between MFs and external fields, and thus only the antisymmetric part  contributes to the coupling:
\begin{align}
  \hat{\mathcal{O}}_{\rm MF}  ({\bm x})=  -\frac{1}{8} \sum_{a,b} [\hat{\gamma}_a, \hat{\gamma}_b] \; \tr \left[  \mathcal{O} \rho^{(ab)} ({\bm x}) \right], \label{eq:OMF}
\end{align}
where $\rho^{(ab)}$ is given by
\begin{align}
 \rho^{(ab)}({\bm x}) \equiv   | u_{0}^{(a)} ({\bm x}) \rangle \langle \mathcal{C}u_{0}^{(b)} ({\bm x}) | - | u_{0}^{(b)} ({\bm x}) \rangle \langle \mathcal{C} u_{0}^{(a)} ({\bm x}) | .
\end{align}
Equation~(\ref{eq:OMF}) is our main formula to examine electromagnetic structures of MKPs.

For further analysis, we employ a group theoretical approach.
As we will show shortly,  $\rho^{(ab)}$ is a single-valued representation of $G_0^{\bm k}$
under the action  $\rho^{(ab)}\mapsto {\cal U}_g^{\bm k}\rho^{(ab)}{\cal U}_g^{{\bm k}\dagger}$ for $g\in G_0^{\bm k}$.
It is decomposed into irreps as
\begin{align}
\rho^{(ab)} = \sum_{Ai} \rho^{(A)}_i,
 \label{eq:decomp_rho}
 \end{align}
 where $\rho^{(A)}$ is a single-valued irrep of $G_0^{\bm k}$ with the transformation law
  \begin{align}
 \mathcal{U}^{\bm k}_g \rho^{(A)}_i\mathcal{U}_g^{{\bm k}\dagger} = \sum_{j} \rho^{(A)}_j[\mathcal{D}_g^{A}]_{ji}.  \label{eq:decomp_rho_trans} 
 \end{align}
 Here ${\cal D}_g^A$ is a real orthogonal matrix. 
 We can also decompose ${\cal O}$ as
 \begin{align}
 \mathcal{O} = \sum_{Ai} \mathcal{O}^{(A)}_i, \label{eq:decomp_O}
 \end{align}
 where ${\cal O}^{(A)}_i$ is an operator belonging to a single-valued irrep of $G_0^{\bm k}$,
 \begin{align}
\mathcal{U}^{\bm k}_g \mathcal{O}^{(A)}_i\mathcal{U}_g^{{\bm k}\dagger} = \sum_{j} \mathcal{O}^{(A)}_j [\mathcal{D}_g^{A}]_{ji}.
\end{align}
By substituting Eqs.~(\ref{eq:decomp_rho}) and~(\ref{eq:decomp_O}) into Eq.~(\ref{eq:OMF}),
the trace part is recast into
\begin{align}
 &\sum_{A,B} \tr \left[ \mathcal{O}^{(A)}_i \rho^{(B)}_j \right] 
 \notag\\
 &=\frac{1}{|G_0^{\bm k}|}\sum_{A,B,g\in G_0^{\bm k}} \tr \left[ {\cal U}_g^{\bm k}\mathcal{O}^{(A)}_i{\cal U}_g^{{\bm k}\dagger} {\cal U}_g^{\bm k}
 \rho^{(B)}_j {\cal U}_g^{\bm k \dagger}\right] \notag\\
&= \frac{1}{|G_0^{\bm k}|}\sum_{A,B,k,l,g\in G_0^{\bm k}} \tr \left[ \mathcal{O}^{(A)}_k \rho^{(B)}_l  \right] [\mathcal{D}^{A}_g]_{ki} [\mathcal{D}^{B}_g]_{lj}
\notag\\ 
&=\sum_{A,B,l} \frac{1}{d_A} 
\tr \left[ \mathcal{O}^{(A)}_l \rho^{(B)}_l  \right] \delta_{ij} \delta_{A B}. 
\label{eq:selection_rule}
\end{align}
where we have used the orthogonality of irreps:
\begin{align}
\sum_{g \in G^{\bm k}_0} [\mathcal{D}^{A}_g]_{ki} [\mathcal{D}^{B}_g]_{lj} = \frac{|G^{\bm k}_0|}{d_A} \delta_{AB} \delta_{ij} \delta_{kl},
\end{align}
with $d_{A}$ the dimension of $\mathcal{D}^{A}_g$. 
Therefore, $\hat{\cal O}_{\rm MF}$ in Eq.(\ref{eq:OMF}) is nonzero 
only when $\mathcal{O}$ shares the same irrep as $\rho^{(ab)}$. 
In other words,  the representation of $\rho^{(ab)}$ determines possible $\mathcal{O}$ for MKPs.

Now we explain how to identify the representation of $\rho^{(ab)}$.
A MF $|u_0^{(a)}\rangle$ at a high symmetry point ${\bm k}$ is  a zero mode of $H({\bm k})$ on the 1d subspace $l_{\bm k}$. 
(More specifically, $|u_0^{(a)}\rangle$ is a zero mode of $H^{\alpha}({\bm k})$ on $l_{\bm k}$.)
Since
${\cal U}_g^{\bm k}$ ($g\in G_0^{\bm k}$) commutes with $H({\bm k})$ on $l_{\bm k}$,  
${\cal U}_g^{\bm k}|u_0^{(a)}\rangle$ is also a zero mode, which implies the relation
\begin{align}
{\cal U}_g^{{\bm k}}|u_0^{(a)}  \rangle =\sum_b | u_0^{(b)} \rangle [U_{g, \gamma}]_{ba}, \label{eq:g_action_Majorana}
\end{align}
with $[U_{g,\gamma}]_{ab} \equiv \langle u_0^{(a)}  | \mathcal{U}^{\bm k}_g  | u_0^{(b)} \rangle$.
Here $U_{g,\gamma}$ obeys the same multiplication law as ${\cal U}_g^{\bm k}$, and thus  it is a double-valued projective representation of $G_0^{\bm k}$ with the same factor system as ${\cal U}_g^{\bm k}$. 
Similarly,  the group action on $|\mathcal{C} u_0^{(a)}  \rangle $ is given  by
\begin{align} 
 \mathcal{U}^{\bm k}_g  |\mathcal{C}  u_0^{(a)} \rangle &
 = \eta_g \mathcal{C}\mathcal{U}^{\bm k}_g  |u_0^{(a)} \rangle \notag \\ 
 &= \sum_b \eta_g|\mathcal{C} u_0^{(b)}  \rangle \left( U_{g, \gamma}\right)^{\ast}_{ba},
 \label{eq:action_hole}
\end{align}
where we have used the relation ${\cal U}_g^{\bm k}{\cal U}_C=\eta_g{\cal U}_C({\cal U}_g^{-{\bm k}})^*$ with ${\cal U}_g^{\bm k}={\cal U}_g^{-{\bm k}}$ at a high symmetry point ${\bm k}$.
Therefore, $\rho^{(ab)}$ is transformed as an anti-symmetric product representation of $G_0^{\bm k}$ under the action  $\rho^{(ab)}\mapsto {\cal U}_g^{\bm k}\rho^{(ab)}{\cal U}_g^{{\bm k}\dagger}$ for $g\in G_0^{\bm k}$,
\begin{align}
 \mathcal{U}^{\bm k}_g \rho^{(ab)} \mathcal{U}_g^{{\bm k}\dagger} 
  &= \sum_{c,d} \rho^{(cd)} \eta_{g} [U_{g,\gamma}]_{ca}  [U_{g,\gamma}]_{db}\notag\\
 &\equiv \sum_{c,d}\rho^{(cd)}[\Omega_g]_{(cd)(ab)},
 \label{eq:rep-rhov1}
 \end{align}
 where $\Omega_g$ is given by 
 \begin{align}
 [\Omega_g]_{(cd)(ab)}=
 \frac{\eta_g}{2}  
 \Big(  [U_{g,\gamma}]_{ca}  [U_{g,\gamma}]_{db} -   [U_{g,\gamma}]_{da}  [U_{g, \gamma}]_{cb} \Big).
\end{align}
Since  the left hand side of Eq.(\ref{eq:rep-rhov1}) does not change the sign 
when ${\cal U}_g^{\bm k}\mapsto -{\cal U}_g^{\bm k}$,  
the anti-symmetric product representation is a single-valued representation of $G_0^{\bm k}$.

On the basis of  the standard group theory,  
we can perform the irreducible decomposition of $\rho^{(ab)}$ by calculating the character of $\Omega_{g}$.  
By taking the trace of $\Omega_g$,  the character of the product representation is given by
 \begin{align}
 \chi^{\Omega}_{g} =\frac{\eta_g}{2}  \Big( (\tr \left[ U_{g,\gamma} \right])^2- \tr  \left[ (U_{g,\gamma})^2 \right] \Big). \label{eq:chafull}
 \end{align}
The right hand side of Eq.(\ref{eq:chafull}) is evaluated as follows. 
In general,  MFs at ${\bm k}$ consist of a set of irreps; 
when a MF $|u_0^{(a)}\rangle$ originates from $H^{\alpha}({\bm k})$ on $l_{\bm k}$,   it is the same irrep $\alpha$ as $H^{\alpha}({\bm k})$.
Correspondingly,  $U_{g,\gamma}$ is decomposed into
\begin{align}
U_{g,\gamma}=\oplus_\alpha U_g^{\alpha},
\end{align}
and thus we have
\begin{align}
 \chi_{g}^{\Omega} =
  \frac{\eta_g}{2}  \left( \left( \sum_{\alpha} \tr \left[ U_{g}^{\alpha} \right] \right)^2- \sum_{\alpha} \tr  \left[ (U_g^{\alpha})^2 \right] \right).\label{eq:chamany}
\end{align}
The right hand side of Eq.(\ref{eq:chamany}) can be easily calculated by the character table of irrep $\alpha$, without referring to the explicit form of $U_{g,\gamma}$. 
Finally, comparing $\chi_g^{\Omega}$ with the characters of the single-valued irreps of $G_0^{\bm k}$,  we obtain the irreducible decomposition of $\rho^{(ab)}$.
In Sec.~\ref{sec:application}, we apply  this method to MKPs at the highest symmetry points of all the wallpaper groups.

The distinction between electric and magnetic structures of MKPs can be done by TRS.
Using TRS, ${\cal O}$ and $\rho^{(ab)}$ are decomposed as
\begin{align}
{\cal O}\equiv{\cal O}_++{\cal O}_-,
\quad
\rho^{(ab)}\equiv \rho^{(ab)}_++\rho^{(ab)}_-,
\end{align}
with
\begin{align}
{\cal O}_{\pm}=\frac{{\cal O}\pm {\cal T}{\cal O}^{\dagger}{\cal T}^{-1}}{2},
\quad 
\rho^{(ab)}_{\pm}=\frac{\rho^{(ab)}\pm {\cal T}\rho^{(ab)\dagger}{\cal T}^{-1}}{2},
\label{eq:O+-rho+-}
\end{align}
where ${\cal O}_{\pm}$ and $\rho^{(ab)}_{\pm}$ satisfy
\begin{align}
{\cal T}{\cal O}^{\dagger}_{\pm}{\cal T}^{-1}=\pm{\cal O}_{\pm},
\quad
{\cal T}\rho^{(ab)\dagger}_{\pm}{\cal T}^{-1}=\pm\rho^{(ab)}_{\pm}.
\label{eq:parityTRS}
\end{align}
Since ${\cal O}$ is hermitian,  the ${\cal O}_{\pm}$ component of ${\cal O}$ is nothing but the even/odd parity component of ${\cal O}$ under TRS.
For TRS,  it holds that
\begin{align}
{\rm tr}\left[{\cal O}\rho^{(ab)}\right]&={\rm tr}\left[{\cal T}({\cal O}\rho^{(ab)})^{\dagger}{\cal T}^{-1}\right]
\nonumber\\
&={\rm tr}\left[{\cal T}\rho^{(ab)\dagger}{\cal T}^{-1}{\cal T}{\cal O}^{\dagger}{\cal T}^{-1}\right]
\nonumber\\
&={\rm tr}\left[
{\cal T}{\cal O}^{\dagger}{\cal T}^{-1}
{\cal T}\rho^{(ab)\dagger}{\cal T}^{-1}
\right],
\label{eq:TRSrhoO}
\end{align}
and thus we obtain 
\begin{align}
{\rm tr}\left[{\cal O}_+\rho^{(ab)}_-\right]
={\rm tr}\left[{\cal O}_-\rho^{(ab)}_+\right]=0.
\end{align}
Therefore, the trace part of Eq.(\ref{eq:OMF}) is given by
\begin{align}
{\rm tr}\left[{\cal O}\rho^{(ab)}\right]
={\rm tr}\left[{\cal O}_+\rho^{(ab)}_+\right]
+{\rm tr}\left[{\cal O}_-\rho^{(ab)}_-\right].
\label{eq:TRS++--}
\end{align}
As we will show in Sec.\ref{sec:emcoupling},  the ${\cal O}_+$ (${\cal O}_-$) component gives the primary coupling to electric (magnetic) fields because an electric (magnetic) field is even (odd) under TRS.
Hence, Eq.(\ref{eq:TRS++--}) means that  the leading coupling  of MKPs to electric (magnetic) fields is determined by $\rho_+^{(ab)}$ ($\rho_-^{(ab)}$).

When the system hosts only a single MKP,  we have additional simplification on $\rho^{(ab)}$.
In this case, $\rho^{(ab)}$ consists of a single component $\rho^{(12)}$ since only two Majorana zero modes exist.
We find that this single component satisfies 
\begin{align}
{\cal T}\rho^{(12)\dagger}{\cal T}=-\rho^{(12)},
\label{eq:Toddrho}
\end{align}
as is shown in the following.
Because the two Majorana zero modes $|u_0^{(a)}\rangle$ $(a=1,2)$ are related by TRS,  we can rewrite $\rho^{(12)}$ as
 \begin{align}
 \rho^{(12)}=|u_0^{(1)}\rangle\langle {\cal C}{\cal T}u_0^{(1)}|-|{\cal T}u_0^{(1)}\rangle\langle {\cal C}u_0^{(1)}|,
 \end{align}
 which leads to
 \begin{align}
 {\cal T}\rho^{(12)\dagger}{\cal T}^{-1}=-|{\cal C}u_0^{(1)}\rangle\langle {\cal T}u_0^{(1)}|
 +|{\cal T}{\cal C}u_0^{(1)}\rangle\langle u_0^{(1)}|.
 \label{eq:TrhoT}
 \end{align}
 Furthermore,  when $|u_0^{(1)}\rangle$ is protected by $\mathbb{Z}_2$ ($\mathbb{Z}$),  we have  $|{\cal C}u_0^{(1)}\rangle=|u_0^{(1)}\rangle$ ($\Gamma |u_0^{(1)}\rangle=\lambda |u_0^{(1)}\rangle$ with $\Gamma=-i{\cal T}{\cal C}$ and $\lambda=\pm 1$).
 In either case,  these relations give  the right hand side of Eq.(\ref{eq:TrhoT}) as $-\rho^{(12)}$, and thus we obtain Eq.(\ref{eq:Toddrho}).
Equation (\ref{eq:Toddrho}) implies that $\rho^{(12)}=\rho_-^{(12)}$ and thus only ${\cal O}_-$ can couple to a single MKP. 
Thus a magnetic field gives the primary coupling to a MKP. (See also Sec.~\ref{sec:emcoupling}).
We also find that the single component $\rho^{(12)}$ and the gap function share the same irrep in this case.
This property follows from that $U_{g,\gamma}$ for a single MKP is given by a rotation matrix of  a spin $J/2$ fermion. 
Since the antisymmetric product of spin $J/2$ fermions is spin-singlet, $\chi_g^{\Omega}$ in Eq. (\ref{eq:chafull}) is readily calculated as
\begin{align}
 \chi_g ^{\Omega}= \eta_g. \label{eq:corr}
\end{align}
We also confirm this relation for all single MKP cases in Tables \ref{tab:EAZsym} and \ref{tab:EAZnonsym} in Sec. \ref{sec:application}.
Equation (\ref{eq:corr}) implies that the representation of $\rho^{(12)}$ coincides with that of the gap function.

If the system hosts more than a single MKP, both $\rho^{(ab)}_+$ and $\rho^{(ab)}_-$ can be nonzero and they can
have more different irreps than the gap function. 
As is shown in Appendix \ref{app:rep},  
$\rho^{(ab)}_{\pm}$ is transformed as
\begin{align}
{\cal U}_g^{\bm k}\rho_{\pm}^{(ab)}{\cal U}_g^{{\bm k}\dagger}\equiv
\sum_{cd}\rho_{\pm}^{(cd)}[\Omega_g^{\pm}]_{(cd)(ab)},
\label{eq:Omega+-}
\end{align}
where $\Omega_g^\pm$ is given by
\begin{align}
&[\Omega_g^{\pm}]_{(cd)(ab)}
\nonumber\\
&=
 \frac{\eta_g}{4}  
 \Big(  [U_{g,\gamma}]_{ca}  [U_{g,\gamma}]_{db} -   [U_{g,\gamma}]_{da}  [U_{g, \gamma}]_{cb} \Big)
 \nonumber\\
 &\pm
 \frac{\eta_g}{4}  
 \Big(  [\Gamma_\gamma U_{g,\gamma}]_{ca}  [\Gamma_\gamma U_{g,\gamma}]_{db} -   [\Gamma_\gamma U_{g,\gamma}]_{da}  [\Gamma_\gamma U_{g, \gamma}]_{cb} \Big).
  \end{align}
with $[\Gamma_\gamma]_{ab}\equiv\langle u_0^{(a)}|\Gamma|u_0^{(b)}\rangle$.
By taking the trace of $\Omega_g^{\pm}$, the character of the representation reads
\begin{align}
\chi_g^{\Omega^{\pm}}
&=\frac{\eta_g}{4}
\left(
(\tr[(U_{g,\gamma}])^2-\tr [(U_{g,\gamma})^2]
\right)
\nonumber\\
&\pm \frac{\eta_g}{4}
\left(
(\tr[\Gamma_\gamma U_{g,\gamma}])^2-\tr [(\Gamma_\gamma U_{g,\gamma})^2]
\right).
\label{eq:chiOmega+-}
\end{align}
From $\chi_g^{\Omega^{\pm}}$,  we can identify the irreps of $\rho_{\pm}^{(ab)}$.

\subsection{electric and magnetic couplings of MKPs}
\label{sec:emcoupling}
The presence of electric or magnetic fields may induce low energy couplings between these external fields and MKPs.
Here we will explain how to determine  such couplings by symmetry. 

First, we consider possible magnetic couplings of MKPs induced by a magnetic field ${\bm B}$.
A magnetic field ${\bm B}$ induces an effective coupling  $\hat{H}_{\rm m}$ between the quantum field $\Psi({\bm x})$ and ${\bm B}$.
In the low energy limit,  $\hat{H}_m$ does not contain any derivatives of $\Psi({\bm x})$ and ${\bm B}$ in the low energy limit,  and thus it can be written as
\begin{align}
\hat{H}_{\rm m}=\int d{\bm x}g({\bm B})\hat{\cal O}({\bm x}),
\end{align}
where $g({\bm B})$ is a real function of ${\bm B}$, and $\hat{\cal O}$  is a local quantum operator in the form of  Eq.(\ref{eq:q-op}). 
\footnote{Here $g({\bm B})$ and $\hat{\cal O}$ can be multi-component, and the summation of the multi-component indices are implicit.}  
Hence, the coupling between MKPs and the magnetic field is given by
\begin{align} 
\hat{H}_{\rm m}=\int d{\bm x}g({\bm B})\hat{\cal O}_{\rm MF}({\bm x}),
\label{eq:Hem}
\end{align}
with $\hat{\cal O}_{\rm MF}$ in Eq.(\ref{eq:OMF}).

Possible $g({\bm B})$ and $\hat{\cal O}_{\rm MF}$ are subject to symmetry.
In general,  a magnetic field breaks (a part of) symmetry of the system. 
However, if one applies the symmetry operation to the magnetic field as well as the quantum operator $\hat{\Psi}$,  the whole system recovers the symmetry, and thus $\hat{H}_{\rm m}$ should be invariant under this symmetry operation.

Under TRS,  $g({\bm B})$ and $\hat{\cal O}_{\rm MF}$ are transformed as
\begin{align}
g({\bm B})\rightarrow g( -{\bm B}),
\end{align}
and
\begin{align}
\hat{\cal O}_{\rm MF}&\rightarrow 
-\frac{1}{8}\sum_{a,b}\left[\hat{\gamma_a},\hat{\gamma}_b\right]
{\rm tr}[{\cal U}_T^{\dagger}\rho^{(ab)}{\cal U}_T{\cal O}^*]
\nonumber\\
&=-\frac{1}{8}\sum_{a,b}\left[\hat{\gamma_a},\hat{\gamma}_b\right]
{\rm tr}[\rho^{(ab)}{\cal T}{\cal O}{\cal T}^{-1}].
\end{align}
Therefore, if we decompose $g ({\bm B})$ and $\hat{\cal O}_{{\rm MF}}$ as
\begin{align}
&g ({\bm B})=g_{+}({\bm B})+g_{-}({\bm B}), 
\nonumber\\
&\hat{\cal O}_{{\rm MF}}=\hat{\cal O}_{{\rm MF}+}+\hat{\cal O}_{{\rm MF}-},
\end{align}
where $g_{+}({\bm B})$ ($g_{-}({\bm B})$) is an even (odd) function of ${\bm B}$,  and 
$\hat{\cal O}_{{\rm MF}\pm}$ is given by 
\begin{align}
\hat{\cal O}_{{\rm MF}\pm}=-\frac{1}{8}\sum_{a,b}\left[\hat{\gamma_a},\hat{\gamma}_b\right]
{\rm tr}[\rho^{(ab)}_\pm{\cal O}_\pm],
\end{align}
then TRS leads to
\begin{align}
g_+({\bm B})\hat{\cal O}_{{\rm MF}-}=g_-({\bm B})\hat{\cal O}_{{\rm MF}+}=0.
\end{align}
Thus, we have 
\begin{align}
\hat{H}_{\rm m}=\int d{\bm x} \left(g_+({\bm B})\hat{\cal O}_{{\rm MF}+}({\bm x})+g_-({\bm B})\hat{\cal O}_{{\rm MF}-}({\bm x})\right).
\label{eq:f+f-}
\end{align}

A further constraint is obtained by crystalline symmetry. 
For $g=\{p|{\bm a}_p\}\in G_0^{\bm k}$, $g_{\pm}({\bm B})$ and $\hat{\cal O}_{{\rm MF}\pm}$ are transformed as
\begin{align}
&g_\pm({\bm B})\rightarrow g_\pm(({\rm det}p)p {\bm B}),
\label{eq:fg}
\\
&\hat{\cal O}_{{\rm MF}\pm}\rightarrow -\frac{1}{8}\sum_{a,b}\left[\hat{\gamma_a},\hat{\gamma}_b\right]
{\rm tr}[{\cal U}_g^{{\bm k}\dagger}\rho_\pm^{(ab)}{\cal U}_g^{\bm k}{\cal O}_{\pm}].
\label{eq:Og}
\end{align}
In order that $\hat{H}_{\rm m}$ is invariant under $G_0^{\bm k}$,  $g_+({\bm B})$ ($g_-({\bm B})$) 
and $\hat{O}_{{\rm MF}+}$ ($\hat{O}_{{\rm MF}-}$) 
should be the same irrep under the transformation in Eqs.(\ref{eq:fg}) and (\ref{eq:Og}).

In a similar manner, we can obtain possible electric couplings of MKPs induced by an electric (polarization) field ${\bm E}$.
TRS requires that the electric couplings should has the following form 
\begin{align}
\hat{H}_{\rm e}=\int d{\bm x} f({\bm E})\hat{\cal O}_{{\rm MF}+}({\bm x}),
\label{eq:g}
\end{align}
since ${\bm E}$ is invariant under TRS.
Furthermore, $g({\bm E})$ and $\hat{\cal O}_{{\rm MF}+}({\bm x})$ should be the same irrep under crystalline symmetry define by $f({\bm E})\rightarrow f(p{\bm E})$ and Eq.(\ref{eq:Og}). We note that though applying an electric field to superconductors is difficult, electric responses can be observed via a distortion of the crystal.

As discussed in Sec.\ref{sec:multipole_op}, for a single MKP,  only 
$\hat{\cal O}_{{\rm MF}-}$ is nonzero. 
Therefore, a single MKP may host only the magnetic coupling.
This result is consistent with the fact that TRS protects a single MKP and a time-reversal breaking magnetic field is necessary to gap it out.
We need more than a single MKP to obtain the electric coupling.

\begin{table*}[tbp]
\caption{EAZ symmetry classes, irreps of $\rho^{(ab)}_-$ and magnetic responses for single MKPs for 2d point groups. The first low in each table shows the wallpaper groups, irreps of MFs, and the effective spin of MFs, where groups in parentheses represent the Schoenfies notations. For each table, the first, second, third, forth, and fifth columns show irreps of gap functions, the emergent Altland Zirnbauer classes, the 1d invariants, irreps of $\rho_-^{(ab)}$, and the leading term of $g_-$, respectively. Here, ``IR'' stands for irreps and we adopt the notation of irreps in the Bilbao Crystallographic Server~\cite{Elcoro17}. Note that irreps of MFs are double-valued irreps and irreps of $\Delta$ and $\mathcal{O}$ are single-valued irreps.
} \label{tab:EAZsym}
\begin{center}
\begin{tabular}{cccccp{2mm}ccccc} \hline\hline
 \multicolumn{5}{c}{ $p1$ (C$_1$), $\bar{A} $, spin 1/2}           
 && 
 \multicolumn{5}{c}{ $p2$ (C$_2$), $(^1\bar{E} ,^2\bar{E})$, spin 1/2}  
 \\  \cline{1-5}\cline{7-11}
IR of $\Delta$ & EAZ & 1dim.  & IR of $\rho_-^{(12)}$  & Magetic multipole $g_-$
 &&
IR of $\Delta$ & EAZ & 1dim. & IR of $\rho_-^{(12)}$  &  Magetic multipole $g_-$
\\ \cline{1-5}\cline{7-11}
 A & DIII & $\mathbb{Z}_2$ & A&  $B_x$, $B_y$, $B_z$ 
 && 
 A & AIII &$\mathbb{Z}$ & A& $B_z$
 \\ 
 &&&&&&
 B & D & $\mathbb{Z}_2$ & B & $B_x$, $B_y$
 \\ \\
 \multicolumn{5}{c}{ $p3$ (C$_3$), $(^1\bar{E},^2\bar{E})$, spin 1/2}           
 && 
 \multicolumn{5}{c}{ $p3$ (C$_3$), $\bar{E}$, spin 3/2}  
 \\  \cline{1-5}\cline{7-11}
IR of $\Delta$ & EAZ & 1dim.  & IR of $\rho_-^{(12)}$  &  Magetic multipole $g_-$
 &&
IR of $\Delta$ & EAZ & 1dim. & IR of $\rho_-^{(12)}$  &  Magetic multipole $g_-$
\\ \cline{1-5}\cline{7-11}
A & AIII & $\mathbb{Z}$ & A & $B_z$
&&
A & DIII & $\mathbb{Z}_2$ & A & $B_z$
\\ \\
\multicolumn{5}{c}{ $p4$ (C$_4$), $(^1\bar{E}_1,^2\bar{E}_1)$ or $(^1\bar{E}_2,^2\bar{E}_2)$, spin 1/2 or 3/2 }           
 && 
 \multicolumn{5}{c}{ $p6$ (C$_6$), $(^1\bar{E}_2,^2\bar{E}_2)$ or $(^1\bar{E}_3,^2\bar{E}_3)$, spin 1/2 or 5/2}  
 \\  \cline{1-5}\cline{7-11}
IR of $\Delta$ & EAZ & 1dim.  & IR of $\rho_-^{(12)}$  &  Magetic multipole $g_-$
 &&
IR of $\Delta$ & EAZ & 1dim. & IR of $\rho_-^{(12)}$  &  Magetic multipole $g_-$
\\ \cline{1-5}\cline{7-11}
A & AIII & $\mathbb{Z}$ & A & $B_z$
&&
A & AIII & $\mathbb{Z}$ & A & $B_z$
\\
B & A & 0 & $-$ &$-$
&&
B & A & 0 & $-$ &$-$
\\ \\
\multicolumn{5}{c}{ $p6$ (C$_6$), $(^1\bar{E}_1,^2\bar{E}_1)$, spin 3/2 }           
 && 
 \multicolumn{5}{c}{ $pm$ (C$_s$), $(^1\bar{E},^2\bar{E})$, spin 1/2 }  
 \\  \cline{1-5}\cline{7-11}
IR of $\Delta$ & EAZ & 1dim.  & IR of $\rho_-^{(12)}$  &  Magetic multipole $g_-$
 &&
IR of $\Delta$ & EAZ & 1dim. & IR of $\rho_-^{(12)}$  &  Magetic multipole $g_-$
\\ \cline{1-5}\cline{7-11} 
A & AIII & $\mathbb{Z}$ & A & $B_z$
&&
A & AIII & $\mathbb{Z}$ & A& $B_z$
\\
B & D & $\mathbb{Z}_2$ & B & $B_x^2-3B_xB_y^2$, $B_y^3-3B_yB_x^2$
&&
B & D & $\mathbb{Z}_2$ & B & $B_x$, $B_y$
\\ \\
\multicolumn{5}{c}{ $pmm$ (C$_{2v}$), $\bar{E}$, spin 1/2  }           
 && 
 \multicolumn{5}{c}{ $p31m$, $p3m1$ (C$_{3v}$), $\bar{E}_1$, spin 1/2 }  
 \\  \cline{1-5}\cline{7-11}
IR of $\Delta$ & EAZ & 1dim.  & IR of $\rho_-^{(12)}$  &  Magetic multipole $g_-$
 &&
IR of $\Delta$ & EAZ & 1dim. & IR of $\rho_-^{(12)}$  &  Magetic multipole $g_-$
\\ \cline{1-5}\cline{7-11} 
A$_1$ & CI & 0 & $-$& $-$
&&
A$_1$ & CI & 0 & $-$ & $-$
\\
A$_2$ & BDI & $\mathbb{Z}$ & A$_2$ & $B_z$
&&
A$_2$ & BDI & $\mathbb{Z}$ & A$_2$ & $B_z$
\\
B$_1$ & BDI & $\mathbb{Z}$ & B$_1$ & $B_y$
&&
&&&&
\\
B$_2$ & BDI & $\mathbb{Z}$ & B$_2$ & $B_x$
&&
&&&&
\\ \\
\multicolumn{5}{c}{ $p31m$, $p3m1$ (C$_{3v}$), $\bar{E}$, spin 3/2  }           
 && 
 \multicolumn{5}{c}{ $p4m$ (C$_{4v}$), $\bar{E}_1$ or $\bar{E}_2$, spin 1/2 or 3/2 }  
 \\  \cline{1-5}\cline{7-11}
IR of $\Delta$ & EAZ & 1dim.  & IR of $\rho_-^{(12)}$  & Magetic multipole $g_-$
 &&
IR of $\Delta$ & EAZ & 1dim. & IR of $\rho_-^{(12)}$  & Magetic multipole $g_-$
\\ \cline{1-5}\cline{7-11} 
A$_1$ & AIII & $\mathbb{Z}$ & A$_1$ & $B_x^3-3B_x B_y^2$
&&
A$_1$& CI & 0 & $-$& $-$
\\
A$_2$ & D& $\mathbb{Z}_2$ & A$_2$ & $B_z$
&&
A$_2$ & BDI & $\mathbb{Z}$ & A$_2$ & $B_z$
\\
&&&&
&&
B$_1$ & AI & 0 & $-$ & $-$
\\
&&&&
&&
B$_2$ & AI & 0 & $-$ & $-$
\\ \\
 \multicolumn{5}{c}{ $p6m$ (C$_{6v}$), $\bar{E}_1$ or $\bar{E}_2$, spin 1/2 or 5/2 }  
 &&
 \multicolumn{5}{c}{ $p6m$ (C$_{6v}$), $\bar{E}_3$, spin 3/2 }  
 \\  \cline{1-5}\cline{7-11}
IR of $\Delta$ & EAZ & 1dim.  & IR of $\rho_-^{(12)}$  & Magetic multipole $g_-$
 &&
IR of $\Delta$ & EAZ & 1dim. & IR of $\rho_-^{(12)}$  & Magetic multipole $g_-$
\\ \cline{1-5}\cline{7-11} 
A$_1$ & CI & 0 & $-$ & $-$
&&
A$_1$ & CI & 0 & $-$ & $-$
\\
A$_2$ & BDI & $\mathbb{Z}$ & A$_2$ & $B_z$
&& 
A$_2$ & BDI & $\mathbb{Z}$ & A$_2$ &  $B_z$
\\
B$_1$ & AI & 0 & $-$ & $-$
&&
B$_1$ & BDI & $\mathbb{Z}$ & B$_1$ & $B_y^3-3B_yB_x^2$
\\
B$_2$ & AI & 0 & $-$ & $-$
&&
B$_2$ & BDI & $\mathbb{Z}$ & B$_2$ & $B_x^3-3B_x B_y^2$
\\
\hline\hline
 \end{tabular}
\end{center}
\end{table*}

\begin{table*}[tbp]
\caption{EAZ symmetry classes, irreps of $\rho^{(ab)}_-$, and magnetic responses for single MKPs for other wallpaper groups including nonsymmorphic ones. Those SGs are explicitly defined by $pg =\{ \{e|0\}, \{ \sigma_y|\bm{\tau}_x \} \}$, $cm =\{\{e|0\}, \{\sigma_{(010)}|0\},\{e|\bm{\tau}_x \}\}$, $pmg =\{\{e|0\}, \{2_z|0\},\{\sigma_{(010)}|\bm{\tau}_x\}, \{\sigma_{(100)}|\bm
{\tau}_x\} \}$, $pgg =\{\{e|0\}, \{2_z|0\}, \{\sigma_{(010)}|\bm{\tau}_x+\bm{\tau}_y\}, \{\sigma_{(100)}|\bm{\tau}_x+\bm{\tau}_y\} \}$, $cmm =\{\{e|0\}, \{2_z|0\}, \{\sigma_{(010)}|0\}, \{\sigma_{(100)}|0\}, \{e|\bm{\tau}_x+\bm{\tau}_y\}  \}$, $p4g =\{\{e|0\}, \{2_z|0\}, \{4_z|0\},  \{\sigma_{(010)}|\bm{\tau}_x+\bm{\tau}_y\}, \{\sigma_{(100)}|\bm{\tau}_x+\bm{\tau}_y\}, \{\sigma_{(110)}|\bm{\tau}_x+\bm{\tau}_y\}, \{\sigma_{(1\bar{1}0)}|\bm{\tau}_x+\bm{\tau}_y\}  \}$, where $\bm{\tau}_i$ is a half translation along the $i$ direction, $n_z$ a $n$-fold rotation around the $z$ axis, $\sigma_{(hkl)}$ a mirror reflection in terms of the ($hkl$) plane. $\bar{\Gamma}$, $\bar{X}$, and $\bar{M}$ label $(0, 0)$, $(\pi, 0)$, and $(\pi, \pi)$ points in the surface BZ. We use the notation of irreps in the Bilbao Crystallographic Server~\cite{Elcoro17} when irreps are similar to those in symmorphic groups. On the other hand, irreps of nonsymmorphic groups at the BZ boundary, $X_i$ and $X_i'$, are given in Table~\ref{tab:nonsym}.
} \label{tab:EAZnonsym}
\begin{center}
\begin{tabular}{cccccp{2mm}ccccc} \hline\hline
 \multicolumn{5}{c}{ $pg$ (C$_s$) $\bar{\Gamma}$ point, $(^1\bar{E}, ^2\bar{E})$, spin 1/2}           
 && 
 \multicolumn{5}{c}{ $pg$ (C$_2$) $\bar{X}$ point, $(X_1, X_1)$ or $(X_2, X_2)$, spin 1/2}  
 \\  \cline{1-5}\cline{7-11}
IR of $\Delta$ & EAZ & 1dim.  & IR of $\rho_-^{(12)}$  & Magnetic multipole $g_-$
 &&
IR of $\Delta$ & EAZ & 1dim. & IR of $\rho_-^{(12)}$  & Magnetic multipole $g_-$
\\ \cline{1-5}\cline{7-11}
 A & AIII & $\mathbb{Z}$ & A& $B_z$
 && 
 A & DIII &$\mathbb{Z}_2$ & A& $B_z$
 \\ 
 B & D & $\mathbb{Z}_2$ & B & $B_x$, $B_y$
 &&
 B & AII & 0 &$-$ & $-$
 \\ \\
 \multicolumn{5}{c}{ $cm$ (C$_s$) $\bar{\Gamma}$ point, $(^1\bar{E},^2\bar{E})$, spin 1/2}           
 && 
 \multicolumn{5}{c}{ $pmg$ (C$_{2v}$) $\bar{\Gamma}$ point, $\bar{E}$, spin 1/2}  
 \\  \cline{1-5}\cline{7-11}
IR of $\Delta$ & EAZ & 1dim.  & IR of $\rho_-^{(12)}$  & Magnetic multipole $g_-$
 &&
IR of $\Delta$ & EAZ & 1dim. & IR of $\rho_-^{(12)}$  & Magnetic multipole $g_-$
\\ \cline{1-5}\cline{7-11}
A & AIII & $\mathbb{Z}$ & A & $B_z$
&&
A$_1$ & CI & 0 & $-$ & $-$
\\
B & D & $\mathbb{Z}_2$& B & $B_x$, $B_y$
&&
A$_2$ & BDI & $\mathbb{Z}$ & A$_2$ & $B_z$
\\
&&&&
&&
B$_1$ & BDI & $\mathbb{Z}$ & B$_1$ & $B_y$
\\
&&&&
&&
B$_2$ & BDI & $\mathbb{Z}$ & B$_2$ & $B_x$
\\ \\
\multicolumn{5}{c}{ $pmg$ (C$_{2v}$) $\bar{X}$ point, $(X'_1, X'_2)$ or $(X_3', X_4')$, spin 1/2 }           
 && 
 \multicolumn{5}{c}{ $cmm$ (C$_{2v}$) $\bar{\Gamma}$ point, $\bar{E}$, spin 1/2}  
 \\  \cline{1-5}\cline{7-11}
IR of $\Delta$ & EAZ & 1dim.  & IR of $\rho_-^{(12)}$  & Magnetic multipole $g_-$
 &&
IR of $\Delta$ & EAZ & 1dim. & IR of $\rho_-^{(12)}$  & Magnetic multipole $g_-$
\\ \cline{1-5}\cline{7-11}
A$_1$ & AIII & $\mathbb{Z}$ & A$_1$ & $B_xB_yB_z$
&&
A$_1$ & CI & 0 & $-$ & $-$
\\
A$_2$ & A & 0 & $-$ & $-$
&&
A$_2$ & BDI & $\mathbb{Z}$ & A$_2$ & $B_z$
\\
B$_1$ & D & $\mathbb{Z}_2$ & B$_1$ & $B_y$
&&
B$_1$ & BDI & $\mathbb{Z}$ & B$_1$ & $B_y$
\\
B$_2$ & A & 0 & $-$ &$-$
&&
B$_2$ & BDI & $\mathbb{Z}$ & B$_2$ & $B_x$
\\ \\
\multicolumn{5}{c}{ $pgg$ (C$_{2v}$) $\bar{\Gamma}$ point, $\bar{E}$, spin 1/2  }           
 && 
 \multicolumn{5}{c}{ $pgg$ (C$_{2v}$) $\bar{X}$ point, $(X_1',X_2')$ or $(X_3', X_4')$, spin 1/2 }  
 \\  \cline{1-5}\cline{7-11}
IR of $\Delta$ & EAZ & 1dim.  & IR of $\rho_-^{(12)}$  & Magnetic multipole $g_-$
 &&
IR of $\Delta$ & EAZ & 1dim. & IR of $\rho_-^{(12)}$  & Magnetic multipole $g_-$
\\ \cline{1-5}\cline{7-11} 
A$_1$ & CI & 0 & $-$ & $-$
&&
A$_1$ & AIII & $\mathbb{Z}$ & A$_1$& $B_xB_yB_z$
\\
A$_2$ & BDI & $\mathbb{Z}$ & A$_2$ & $B_z$
&&
A$_2$ & A & 0 & $-$ & $-$
\\
B$_1$ & BDI & $\mathbb{Z}$ & B$_1$ & $B_y$
&&
B$_1$ & D & $\mathbb{Z}_2$ & B$_1$ & $B_y$
\\
B$_2$ & BDI & $\mathbb{Z}$ & B$_2$ & $B_x$
&&
B$_2$ & A & 0 & $-$ &$-$
\\ \\
\multicolumn{5}{c}{ $p4g$ (C$_{4v}$) $\bar{\Gamma}$ point, $\bar{E}_1$ or $\bar{E}_2$, spin 1/2 or 3/2 }           
 && 
 \multicolumn{5}{c}{ }  
 \\  \cline{1-5}
IR of $\Delta$ & EAZ & 1dim.  & IR of $\rho_-^{(12)}$  & Magnetic multipole $g_-$
 &&
& & &  & 
\\ \cline{1-5} 
A$_1$ & CI & 0 & $-$& $-$
&&
& & & &
\\
A$_2$ & BDI & $\mathbb{Z}$ & A$_2$ & $B_z$ 
&&
& & & &
\\
B$_1$ & AI & 0 & $-$ & $-$
&&
&&&&
\\
B$_2$ & AI & 0 & $-$ & $-$
&&
&&&&
\\
\hline\hline
 \end{tabular}
\end{center}
\end{table*}

\section{Application to the wallpaper groups}
\label{sec:application}
\subsection{Majorana multipole response}
\label{sec:multipole_resp}
We now apply the general theory developed in the previous section to MKPs protected by wallpaper groups. 
We consider the minimal set of MKPs positioned at each of the highest symmetry points in the surface BZ where the little group $G_0^{\bm k}$ is $G_0$ itself. 
The minimal MKPs are systematically determined from the Wignar's test for TRS. As discussed in Sec.\ref{sec:degeneracy}, when $W_{\alpha}^T=1 (-1)$, a Kramers pair is formed in a single (a pair of) $\alpha$, while when  $W_{\alpha}^T=0$, a Kramers pair is formed between different irreps~\cite{Wigner59}. 
The resultant minimal MKPs are listed in Tables~\ref{tab:EAZsym} and~\ref{tab:EAZnonsym}.  
In most cases, the minimal set is  a single MKP, and thus $\rho^{(ab)}$ consists of a single component $\rho_-^{(12)}$.
As is shown in Sec.\ref{sec:multipole_op}, the irrep of $\rho_-^{(12)}$ coincides with that of the gap function. 
For instance, let us consider $p2=\{\{e|\bm{0}\}, \{2_z|\bm{0}\}\}$. In this case, $U_{g,\gamma}$ is given by 
\begin{align}
U_{\{E|\bm{0}\},\gamma} = \begin{pmatrix} 1 & 0 \\ 0& 1 \end{pmatrix}, \quad
U_{\{2_z|\bm{0}\},\gamma} = \begin{pmatrix} i & 0 \\ 0& -i \end{pmatrix}. 
\label{eq:repC2}
\end{align}
Substituting Eq.~(\ref{eq:repC2}) into Eq.~(\ref{eq:chafull}), we find 
\begin{align}
\chi_{\{e|\bm{0}\}}^{\Omega} = 1, \quad \chi_{\{2_z|\bm{0}\}}^{\Omega} = \eta_{2_z},
\end{align}
which reproduces Eq.(\ref{eq:corr}).
We summarize the obtained irreps of $\rho^{(ab)}$ in Tables~\ref{tab:EAZsym} and~\ref{tab:EAZnonsym}.
We note that the minimal set  is double MKPs when MFs are positioned at the $\bar{M}$ point of  $pgg$ or $p4g$. As explained in Sec.\ref{sec:degeneracy}, the double MKPs originate from a crystalline symmetry-enforced fourfold degeneracy. 
In the following, we focus on magnetic responses in the single MKP cases. 
The electromagnetic responses for double MKPs will be discussed in Sec.~\ref{sec:multiple}. 

For a single MKP, a nonzero $\hat{\cal O}_{\rm MF}$ is always odd under TRS (see arguments in Sec.\ref{sec:multipole_op}).
Hence, the magnetic coupling is given by
\begin{align}
\hat{H}_{\rm m}=\int d{\bm x} g_-({\bm B})\hat{\cal O}_{\rm MF-}({\bm x}),
\end{align}
where 
\begin{align}
g_-(\bm{B}) = \sum_i c_{1,i} B_i + \sum_{ijk} c_{3,ijk} B_i B_j B_k+\cdots.
\end{align} 
Here $c_{1,i}$ and $c_{3, ijk}$ are material dependent parameters
subject to constraints from crystalline symmetry.
The symmetry-adopted forms of $g_-(\bm{B})$ are listed in Table~\ref{tab:magrep}, where the first, third, fifth, and seventh order of magnetic fields correspond to a magnetic dipole, octupole, 32-pole, and 128-pole, respectively. 
As argued in Sec.\ref{sec:emcoupling}, $g_-({\bm B})$ should be the same irrep as $\rho_-^{(12)}$.

 For instance, let us consider $p2$.  When the irrep of the gap function is A,  the irrep of $\rho_-^{(12)}$ is also A. 
 Thus an allowed ${\cal O}$ satisfies 
\begin{align}
\mathcal{U}_{\{2_z|0\}} \mathcal{O}_{p2-}^{\rm A} \mathcal{U}_{\{2_z|0\}}^{\dagger} = \mathcal{O}_{p2-}^{\rm A},
\quad
 {\cal T}{\cal O}_{p2-}^{\rm A}{\cal T}^{-1}=-{\cal O}_{p2-}^{\rm A},
\end{align}
and $g_-(\bm{B})$ is given by
\begin{align}
g_{p2-}^{\rm A}(\bm{B})  = c_1B_z,
\end{align}
since a magnetic field is transformed as $(B_x,B_y,B_z) \to (-B_x,-B_y,B_z)$ under $\{2_z|\bm{0}\}$.
Hence $\hat{H}_{\rm m}$ reads
\begin{align}
\hat{H}_{\rm m}=c_1\int d{\bm x} B_z \hat{\cal O}^{\rm A}_{{\rm MF} \, p2-},
\end{align}
where $\hat{\cal O}^{\rm A}_{{\rm MF} \, p2-}$ is defined by Eq.(\ref{eq:OMF}) with ${\cal O}={\cal O}_{p2-}^{\rm A}$.
This term provides the magnetic dipole response of the MKP along the rotation axis. 
Such a magnetic response has been known for superfluid $^3$He-B phase~\cite{Chung09,Nagato09,Mizushima12,Mizushima16}, 
and the E$_{1u}$ state of UPt$_3$~\cite{Tsutsumi13}. 
On the other hand, for the B gap function, ${\cal O}$ for the MKP satisfies
\begin{align}
 \mathcal{U}_{\{2_z|0\}} \mathcal{O}_{p2-}^{\rm B} \mathcal{U}_{\{2_2|0\}}^{\dagger}= -\mathcal{O}_{p2-}^{\rm B}.
\end{align}
Thus, the lowest order of $g_-(\bm{B})$ is given by
\begin{align}
g_{p2-}^{\rm B}(\bm{B}) = c_{1,1} B_x+c_{1,2} B_y,
\end{align}
resulting in the magnetic dipole response parallel to the surface. The direction of the dipole response depends on the material dependent parameters $c_{1,i}$ ($i=1,2$). 

In this manner, we can determine possible magnetic responses of a single MKP protected by all paper groups, which
 are summarized in Tables~\ref{tab:EAZsym} and~\ref{tab:EAZnonsym}. 
 We find that a single MKP shows magnetic dipole responses in most cases: 
 In addition to the $^3$He-B phase and the E$_{1u}$ state of UPt$_3$ mentioned in the above, the A$_{1u}$ state of the superconducting doped topological insulator\cite{Fu10}  also shows the magnetic dipole response.  For the $(111)$ surface of the doped topological insulator, which is normal to the $z$-direction, there exists a single MKP at the $\bar{\Gamma}$ point of 
 the surface BZ. 
 Because the $(111)$ surface hosts $C_{3v}$ symmetry, and the A$_{1u}$ gap function is the A$_2$ irrep for $C_{3v}$, from Table \ref{tab:EAZsym}, we find that the surface MKP has the magnetic dipole parallel to the $z$-direction.

Interestingly, our result shows that magnetic octupole responses are also possible. 
 The rest of this section focuses on the magnetic octupole responses, which are realized in spin 3/2 TSCs and nonsymmorphic TSCs.

\subsection{Magnetic octupole response of spin 3/2 MFs}
\label{sec:high}
The spin of MKPs is effectively given by 1/2, 3/2, and 5/2, each of which forms different irreps. 
There exist magnetic structures allowed only for the spin 3/2 MFs when the wallpaper group including the threefold rotation symmetry.

For example, let us consider the case with $p3m1 = \{\{e|\bm{0}\}, \{3_z| \bm{0} \}, \{\sigma_{(100)} | \bm{0}\}\} $ and the A$_1$ gap function. Here $\{3_z | \bm{0} \}$ is a threefold rotation about the $z$ axis and $\{\sigma_{(100)} | \bm{0} \}$ is a mirror-reflection with respect to the $(100)$ plane. 
 From Table~\ref{tab:EAZsym}, the irrep of $\rho^{(12)}$ is A$_1$. Thus, ${\cal O}_-$ coupled to the spin-$3/2$ MKP satisfies
\begin{subequations}
\begin{align}
 &\mathcal{U}_{\{3_z|\bm{0}\}} \mathcal{O}_{p3m1-}^{{\rm A}_1} \mathcal{U}_{\{3_z|\bm{0}\}}^{\dagger} = \mathcal{O}_{p3m1-}^{{\rm A}_1} , \label{eq:p3m1op1}\\
 &\mathcal{U}_{\{\sigma_{(100)}|\bm{0}\}} \mathcal{O}_{p3m1-}^{{\rm A}_1}  \mathcal{U}_{\{\sigma_{(100)}|\bm{0}\}}^{\dagger} = \mathcal{O}_{p3m1-}^{{\rm A}_1} . \label{eq:p3m1op2}
\end{align}
\end{subequations}
The magnetic field changes as $\{3_z|\bm{0}\}: (B_+, B_-,B_z) \to  ( e^{\frac{i 2\pi}{3}}B_+, e^{-\frac{i 2\pi}{3}} B_- , B_z)$ and $\{\sigma_{(100)}|\bm{0}\}: (B_x,B_y,B_z) \to  (B_x,-B_y,-B_z)$ under the operations of $p3m1$, where $B_+\equiv B_{x} + i B_{y} $, $B_+\equiv B_{x} - i B_{y} $. Since the irrep of $g_-(\bm{B})$ has to be $A_1$, it is of the form:
\begin{align}
 g_{p3m1-}^{\rm A_1}(\bm{B}) =c_{3} \left( B_{x}^3 -3 B_{x} B_{y}^2 \right),
 \label{eq:MRp3m1}
\end{align}
which gives the magnetic octupole response as a leading contribution.
It should be noted here that the three-fold rotation symmetry forbids the first order term of $B_i$. In a similar manner, the magnetic octupole response appears for $p6m$ when the spin of MKPs is $3/2$ and the irrep of the gap functions is the B$_1$ or B$_2$ gap function. The $g_-(\bm{B})$'s in the lowest order are
\begin{align}
  &g_{p6m-}^{\rm B_1}(\bm{B}) =c_{3} \left( B_{y}^3 -3 B_{y} B_{x}^2 \right), \label{eq:MRp6mB1} \\ 
  &g_{p6m-}^{\rm B_2}(\bm{B}) =c_{3} \left( B_{x}^3 -3 B_{x} B_{y}^2 \right). \label{eq:MRp6mB2}
\end{align}

In the previous study~\cite{Kobayashi19}, we pointed out that the magnteic octupole response is realized in the half-Hausler superconductors~\cite{Goll08,Butch11,Tafti13,GXu16,Bay12,Brydon16,Kim18} with the A$_1$ gap function of T$_d$ in the $(111)$ surface. On the surface, the A$_1$ irrep of T$_d$ is compatible with the A$_1$ irrep of C$_{3v}$ ($p31m$ or $p3m1$ in the wallpaper groups). The similar compatible relation is met in the A$_{2u}$ irrep of O$_h$. Hence the antiperovskite Dirac metals~\cite{Oudah16,Kawakami18} with the A$_{2u}$ gap function of O$_h$ are also a possible candidate for this response.

In $p6$ symmetric TSCs with the B gap function, we have a slightly different magnetic octupole response. As is the case with $p3m1$, the spin-3/2 state is necessary, but the MKP is stabilized by a $\mathbb{Z}_2$ invariant. ${\cal O}_-$ for the MKP only respects
\begin{align}
\mathcal{U}_{\{6_z|\bm{0}\}} \mathcal{O}_{p6-}^{\rm B} \mathcal{U}_{\{6_z|\bm{0}\}}^{\dagger} = -\mathcal{O}_{p6-}^{\rm B}. \label{eq:p6op}
\end{align}
Therefore, the magnetic octupole response is described by
\begin{align}
  g_{p6-}^{\rm B} (\bm{B})=c_{3,1} \left( B_{x}^3 -3 B_{x} B_{y}^2 \right) 
  +c_{3,2} \left( B_{y}^3 -3 B_{y} B_{x}^2 \right).
  \label{eq:MRp6}
\end{align}
In the polar coordinate $(B_x,B_y) =B(\cos \phi, \sin \phi)$, it is rewritten as
\begin{align}
g_{p6-}^{\rm B}(\bm{B}) = c_{3,1}B^3 \cos (3 \phi + \theta_{\rho}),
\end{align}
where $\tan \theta_{\rho} = c_{3,2}/c_{3,1}$.  Since the magnetic response preserves only the sixfold rotation symmetry, it can be tilted by the material dependent angle $\theta_{\rho}$. Interestingly, this type of the magnetic ocutpole response is realized in fully-gapped TSCs as we will show in Sec.~\ref{sec:modelp6}.

\subsection{Magnetic octupole response by nonsymmorphic symmetry}
\label{sec:nonsym}
We show another mechanism realizing the magnetic octupole response. The key ingredient is the glide symmetry, which appears in $pg$, $pmg$, $pgg$, and $p4g$. 

For example, we consider the case with $pmg=\{\{e|\bm{0}\}, \{2_z|\bm{0}\}, \{\sigma_{(010)}|\bm{\tau}_x\} \}$ and the A$_1$ gap function.  The irrep of ${\cal O}$ for the MKP at $\bar{\Gamma}$ point is different from that at $\bar{X}$ point since the factor system for glide $\{\sigma_{(010)}|{\bm \tau}_x\}$ has an additional phase at the $\bar{X}$ point. 
From Table~\ref{tab:EAZnonsym}, we have a $\mathbb{Z}$ invariant at the $\bar{X}$ point 
while there is no topological invariant at the $\bar{\Gamma}$ point. When the irrep of ${\cal O}_-$ is A$_1$, it satisfies 
\begin{subequations}
\begin{align}
 &\mathcal{U}_{\{2_z|\bm{0}\}} \mathcal{O}_{pmg-}^{\rm A_1} \mathcal{U}_{\{2_z|\bm{0}\}}^{\dagger}  = \mathcal{O}_{pmg-}^{\rm A_1} , \label{eq:pmgop1}\\
 &\mathcal{U}_{\{\sigma_y|\bm{\tau}_x\}} \mathcal{O}_{pmg-}^{\rm A_1}  \mathcal{U}_{\{\sigma_y|\bm{\tau}_x\}}^{\dagger}  = \mathcal{O}_{pmg-}^{\rm A_1} . \label{eq:pmgop2}
\end{align}
\end{subequations}
On the other hand,  the magnetic field changes as
\begin{subequations}
\begin{align}
&\{2_z|\bm{0}\}: (B_x,B_y,B_z) \to (-B_x,-B_y,B_z), \\
&\{\sigma_y|\bm{\tau}_x\}: (B_x,B_y,B_z) \to (-B_x,B_y,-B_z).
\end{align} 
\end{subequations}
As $g_-({\bm B})$ for the magnetic coupling should be the same irrep as ${\cal O}_-$,  it is given by 
\begin{align}
  g_{pmg-}^{\rm A_1} (\bm{B}) =c_{3} B_x B_y B_z,
  \label{eq:MRpmg}
\end{align}
which represents the magnetic octupole response.

\section{Majorana octupole responses in tight-binding models }
\label{sec:model}

In this section, using concrete models, we demonstrate magnetic octupole responses in $p6$ and $pmg$,  which have been overlooked so far. 

\subsection{Model with $p6$ symmetry}
\label{sec:modelp6}
\begin{figure}[tbp]
\centering
 \includegraphics[width=8cm]{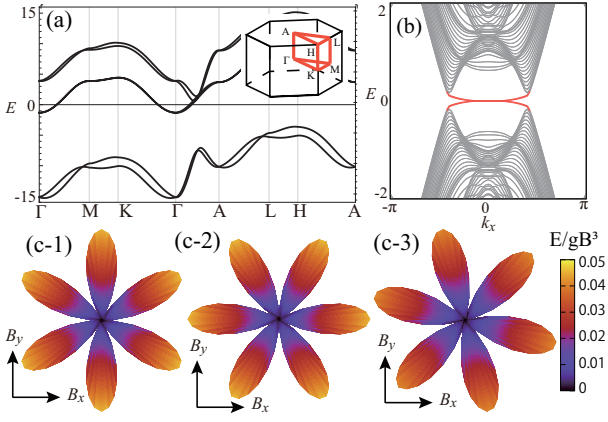}
 \caption{(a) The band structure of Eq.~(\ref{eq:p6model}) with parameters $(m_0,m_1,t_{xy},t_z,\alpha,\beta,\gamma)=(2.2,5,-1.3,-2.5,6,0.5,1)$. (b) The $(001)$ surface state in the fully-gapped superconducting states. Here, we choose the chemical potential and the amplitude of the gap funcion as $\mu=0$ and $\Delta_0=1$ and the B$_1$ and B$_2$ gap functions coexist such that $\eta_1=\eta_2=\eta_1'=\eta_2'=0.5$. (c) Applying the Zeeman magnetic field $h_{\rm Z} = g  \bm{B} \cdot \bm{s}$, the energy gap of the MKP is illustrated as a function of $\bm{B}$: (c-1) $\Delta_{B_1}$ only, (c-2) $\Delta_{B_2}$ only, and (c-3) the mixture of $\Delta_{B_1}$ and $\Delta_{B_2}$.}\label{fig:p6model}
\end{figure}

We consider a tight-binding model with space group $P622$ (SG$\#$ 177), which is built on the triangular lattice with $p_x$, $p_y$, and $p_z$ orbitals on each site. The normal Hamiltonian is given by
\begin{align}
 h_{p6} (\bm{k}) & = h_{0} (\bm{k}) + h_{\rm soc} (\bm{k}),  \label{eq:p6model}
\end{align}
with
\begin{subequations}
\label{eq:p6model_comps}
\begin{align}
h_{0} (\bm{k}) &=m_0 + m_1 \lambda_8 + t_z \cos k_z \notag \\
& +t_{xy} \left\{ \cos k_x + 2 \cos \left(\frac{k_x}{2}\right)  \cos \left(\frac{\sqrt{3}k_y}{2}\right)\right\},  \\
h_{\rm soc} (\bm{k}) &= \alpha \cos k_z (\lambda_5 s_y - \lambda_7 s_x)+\beta \sin k_z (\lambda_4 s_x + \lambda_6 s_y) \notag \\
 & + \gamma (\sin k_x  s_x + \sin k_y  s_y), 
\end{align}
\end{subequations}
where $\lambda_i$ ($i=1-8$) are the Gell-Mann matrices acting on the $(p_x,p_y,p_z)$ orbitals, and $s_i$ are the Pauli matrices in the spin space. $m_0$ and $m_1$ are on-site potentials, $t_z$ and $t_{xy}$ are hopping terms, and $\alpha$, $\beta$, and $\gamma$ represent spin-orbital interactions.
The normal Hamiltonian hosts TRS and the $P662$ symmetry below,
\begin{subequations}
\begin{align}
 U_{\{6_z|\bm{0}\}}h_{p6} (\bm{k}) U_{\{6_z|\bm{0}\}}^{\dagger}  = h_{p6} (6_z \bm{k}), \\
U_{\{2_y|\bm{0}\}} h_{p6} (\bm{k}) U_{\{2_y|\bm{0}\}}^{\dagger} = h_{p6} (2_y \bm{k}), \\
U_{\{2_x|\bm{0}\}} h_{p6} (\bm{k}) U_{\{2_x|\bm{0}\}}^{\dagger} = h_{p6} (2_x \bm{k}),
\end{align}
\end{subequations}
with
\begin{subequations}
\begin{align}
 U_{\{6_z|\bm{0}\}} &= R_z \left(\frac{2\pi}{6} \right) \exp \left ( -i s_z \frac{\pi}{6}\right), \\
 U_{\{2_y|\bm{0}\}} &= R_y (\pi)(-i s_y), \\
 U_{\{2_x|\bm{0}\}} &= R_x (\pi)(-i s_x),
\end{align}
\end{subequations}
where $R_i(\theta)$ is a $3 \times 3$ rotation matrix in the basis $(p_x,p_y,p_z)$ and represents the $\theta$ rotation about the $i$ axis. 
The band structure of Eq.~(\ref{eq:p6model}) is shown in Fig.~\ref{fig:p6model} (a). When $\beta=\gamma=0$, the normal Hamiltonian recovers the spatial inversion, giving three doubly degenerate bands: one band is effectively described by spin $3/2$ electrons, whereas the other bands are described by spin $1/2$ electrons. When $\beta \neq 0$ and $\gamma \neq 0$, the doubly degenerate bands are split due to the breaking of spatial inversion. In the following, we choose the chemical potential such that the spin $3/2$ band forms the Fermi surface around the $\Gamma$ point.  For the superconducting state, we consider the B$_1$ and B$_2$ gap functions:
\begin{subequations}
\begin{align}
 \Delta_{\rm B_1} (\bm{k})&=\Delta_0 \left( \eta_1 \sin k_z  (\lambda_1 s_x + \lambda_3 s_y) + \eta_2 f_x(\bm{k}) \lambda_2 \right) (is_y), \\
 \Delta_{\rm B_2} (\bm{k})&=\Delta_0 \left( \eta_1'\sin k_z  (\lambda_1 s_y - \lambda_3 s_x)+ \eta_2' f_y(\bm{k}) \lambda_2 \right) (is_y) ,
\end{align}
\end{subequations}
where $f_x (\bm{k}) = \left( \sin k_x - 2 \sin \left( \frac{k_x}{2}\right) \cos \left( \frac{\sqrt{3} k_y}{2}\right) \right) $ and $f_y(\bm{k})=\left( \sin k_y - 2 \sin \left( \frac{k_y}{2}\right) \cos \left( \frac{\sqrt{3} k_x}{2}\right) \right)$, and $\eta_1$, $\eta_2$, $\eta_1'$, and $\eta_2'$ are real parameters.  
When the gap function is $\Delta_{\rm B_1}$ ($\Delta_{\rm B_2}$), 
there appear point nodes in the $k_y$ axis (the $k_x$ axis),
which are protected by $\{2_y|\bm{0}\}$ ($\{2_x|\bm{0}\}$) rotation symmetry. 
On the $(001)$ plane, the zero energy flatband states connecting the point nodes appear because a 2d $\mathbb{Z}_2$ invariant becomes non-trivial between the point nodes. 

We here focus on the MKP at the $\bar{\Gamma}$ point $(k_x=k_y=0)$.
Since B$_1$ and B$_2$ in $P622$ are compatible with $B$ in $p6$ on the surface. Hence, our theory predicts the magnetic octupole response there. To demonstrate this, we add the Zeeman magnetic field $h_{\rm Z} = g  \bm{B} \cdot \bm{s}$ in the normal Hamiltonian and numerically calculate the energy gap of the MKP as a function of $\bm{B}$.  See Figs.~\ref{fig:p6model} (c-1) and (c-2). The magnetic responses keep sixfold rotation symmetry and behaves like Eq.~(\ref{eq:MRp3m1}).
The magnetic octupole response is unique to the spin $3/2$ electrons because a point node appears at $k_x=k_y=0$ if  the chemical potential lies on the spin $1/2$ electrons~\cite{Sumita19}.

Furthermore, if the $P622$ symmetry is broken to the $P6$ symmetry, the $\Delta_{\rm B_1}$ and $\Delta_{\rm B_2}$ gap functions can coexist, and the BdG Hamiltonian realizes a fully-gap TSC.  A MKP exists at $k_x=k_y=0$ (see Fig.~\ref{fig:p6model} (b)), which shows a tilted magnetic octupole response as shown in Fig.~\ref{fig:p6model} (c-3). The magnetic response respects the sixfold rotation symmetry, but zeros of the energy gap appear according to Eq.~(\ref{eq:MRp6}) with nonzero $\theta_{\rho}$.

\subsection{Model with $pmg$ symmetry}
\label{sec:modelpmg}

So far, we consider surface MFs in 3d topological superconductors, but our theory also works for MFs in lower dimensional systems. Here we consider a 2d model with $pmg$ symmetry, which host a single MKP showing a magnetic octupole response.

\begin{figure}[tbp]
\centering
 \includegraphics[width=8cm]{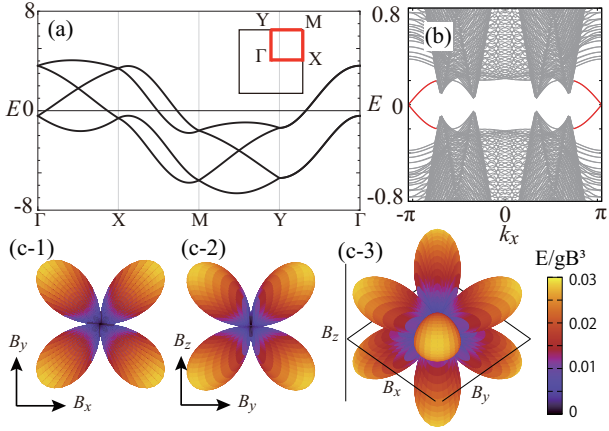}
 \caption{(a) The band structure of Eq.~(\ref{eq:hpmg}) with parameters $(m_0,t_1,t_2,t_3,\alpha,\beta,\gamma)=(-1,0.1,2.5,0.25,-1,0.3,2)$. (b) The $(01)$ surface state in the superconducting state with $\mu = 1$, $\Delta_0=1$, $\eta_1=0.5$, and $\eta_2 = 0.1$. (c) The energy gap of the MKP as a function of $\bm{B}$ under the Zeeman magnetic field $h_{\rm Z} = g \bm{B} \cdot \bm{s}$: The view from (c-1) [001], (c-2) [100], and (c-3) [111] directions.}\label{fig:pmmamodel}
\end{figure}

Let us consider the 2d square lattice with $Pma2$ (SG$\#$ 28)~\cite{QZWang16,Yamazaki19}. In the unit cell, we have two atoms located at $(0,0,-z)$ and $(1/2,0,z)$. Provided that only the $s$ orbital exists on each site, the tight-binding model is   
\begin{align}
h_{pmg}(\bm{k}) =  h_{0}(\bm{k}) + h_{\rm soc}(\bm{k}), \label{eq:hpmg}
\end{align}
with
\begin{subequations}
\begin{align}
h_{0} (\bm{k}) &=m_0 + t_1 \cos k_x + t_2 \cos k_y +t_3 \cos \left( \frac{k_x}{2 }\right)\sigma_x(k_x),  \\
h_{\rm soc} (\bm{k}) &=  (\alpha \sin k_y s_x + \beta \sin k_x s_y )\sigma_z \notag \\
 &+\gamma \left( \sin \left( \frac{k_x}{2} \right) \sigma_x(k_x) s_z + \cos \left( \frac{k_x}{2} \right) \sigma_y(k_x) s_x \right),
\end{align}
\end{subequations}
where $s_i$ and $\sigma_i$ ($i=x,y,z$) are Pauli matrices describing the spin and the sublattice degrees of freedom, and
 $\sigma_x(k_x)$ and $\sigma_y(k_x)$ are modified Pauli matrices:
\begin{subequations}
\begin{align}
&\sigma_x(k_x) \equiv \begin{pmatrix}0 & e^{i k_x/2} \\ e^{-i k_x/2} & 0 \end{pmatrix}, \\
&\sigma_y(k_x) \equiv \begin{pmatrix}0 & -i e^{i k_x/2} \\ ie^{-i k_x/2} & 0 \end{pmatrix}.
\end{align} 
\end{subequations}
$m_0$ is an on-site potential, $t_1$ and $t_2$ are in-plane hopping terms, $t_3$ represents a hopping between the different atoms,  $\alpha$ and $\beta$ are in-plane spin-orbit interactions, and $\gamma$ is spin-orbit interactions between the different atoms. Equation~(\ref{eq:hpmg}) respects TRS and the following crystal symmetries:
\begin{subequations}
\begin{align}
 &U_{\{2_y|\bm{0} \} }h_{pmg}(\bm{k}) U_{\{2_y|\bm{0} \} }^{\dagger} = h_{pmg}(2_y \bm{k}), 
 \label{eq:2y}
 \\
 &U_{\{\sigma_{(001)}|\bm{\tau}_x\}} h_{pmg}(\bm{k}) U_{\{\sigma_{(001)}|\bm{\tau}_x\}}^{\dagger} = h_{pmg}(\sigma_{(001)} \bm{k}),  
 \label{eq:sz}
 \\
 &U_{\{\sigma_{(100)}|\bm{\tau}_x\}} h_{pmg}(\bm{k}) U_{\{\sigma_{(100)}|\bm{\tau}_x\}}^{\dagger} = h_{pmg}(\sigma_{(100)} \bm{k}) ,
\label{eq:sx}
\end{align}
\end{subequations}
with 
\begin{subequations}
\begin{align}
&U_{\{2_y|\bm{0}\}} = -i \sigma_x s_y, \label{eq:pmg2yop}\\
&U_{\{\sigma_{(001)}|\bm{\tau}_x\}} = \begin{pmatrix} 0& e^{ik_x} \\  1 & 0 \end{pmatrix} (i s_z), \label{eq:pmgmzop}\\ 
&U_{\{\sigma_{(100)}|\bm{\tau}_x\}} = \begin{pmatrix} e^{ik_x} & 0 \\ 0 & 1 \end{pmatrix} (i s_x). \label{eq:pmgmxop}
\end{align}
\end{subequations}
We show the band structure of Eq.~(\ref{eq:hpmg}) in Fig.~\ref{fig:pmmamodel} (a), where the glide symmetry-protected band crossing appears on the lines $\Gamma-X$ and $Y-M$. For the superconducting state, we consider the A$_1$ gap function:
\begin{align}
 \Delta_{A_1} = \Delta_0 \left( \eta_1 \sin k_y \sigma_z s_x +\eta_2 \sin \left(\frac{k_x}{2}\right) \sigma_x(k_x) s_z \right)(i s_y).
\end{align}
Numerically diagonalizing the BdG Hamiltonian with the open boundary condition in the $y$ direction, we obtain a single MKP at $k_x=\pi$ as shown in Fig.~\ref{fig:pmmamodel} (b).  
Whereas the open boundary condition breaks the two-fold rotation symmetry in Eq.(\ref{eq:2y}), it keeps $pmg$ symmetry generated by Eqs. (\ref{eq:sz}) and (\ref{eq:sx}).
Then, regarding the $k_x=\pi$ point as the $\bar{X}$ point (and exchanging the $y$ and the $z$ directions in Table \ref{tab:EAZnonsym}), we can use the result in Table \ref{tab:EAZnonsym}, which predicts the magnetic octupole response in the form of Eq.(\ref{eq:MRpmg}).
In Fig.~\ref{fig:pmmamodel} (c), we show the magnetic response obtained by
adding the Zeeman magnetic term $h_{\rm Z} = g \bm{B} \cdot \bm{s}$ in the normal Hamiltonian.
This result is consistent with Eq.~(\ref{eq:MRpmg}).

\section{MKPs at $\bar{M}$ point in $pgg$ and $p4g$}
\label{sec:multiple}

So far, we have considered magnetic responses of a single MKP, which is valid except for the wallpaper groups $pgg$ and $p4g$. For the $pgg$ and $p4g$ symmetries at the $\bar{M}$ point, TRS and crystalline symmetry ensure a fourfold degenerate band crossing on surfaces, realizing two MKPs.   In the following, we discuss electromagnetic structures arising from those two MKPs. 

When there are two MKPs, both $\rho_+^{(ab)}$ and $\rho_-^{(ab)}$ are nonzero, and thus both electric and magnetic couplings are possible. 
Using Eq.(\ref{eq:chiOmega+-}), we determine the electric and magnetic couplings.

\subsection{short representation}

\begin{table*}
 \caption{Short representation of double MKPs at the $\bar{M}$ point for $pgg$ and $p4g$. Matrix representations of other elements in $p4g$ are given by the product of them, e.g., $U_{\{\sigma_{(110)}|{\bm \tau}_x+{\bm \tau}_y\},\gamma}=U_{\{\sigma_{(100)}|{\bm \tau}_x+{\bm \tau}_y\},\gamma} U_{\{4_z|{\bm 0}\},\gamma}$.}
 \label{tab:pgg-p4g-shortrep}
 \begin{center}
 \begin{tabular}{cccccccc} \hline\hline
 \multicolumn{8}{c}{ $pgg$ $\bar{M}$ point}  \\ \cline{1-8}
 IR of $\Delta$ & 
 $U_{\{e|{\bm 0}\},\gamma}$ &
 $U_{\{2_z|{\bm 0}\},\gamma}$ &
 $U_{\{\sigma_{(010)}|{\bm \tau}_x+{\tau}_y\},\gamma}$ &
 $U_{\{\sigma_{(100)}|{\bm \tau}_x+{\bm \tau}_y\},\gamma}$&
 $T_\gamma$ & $C_\gamma$  &$\Gamma_\gamma$\\ \hline
 $A_1$ &$\sigma_0s_0$ & $i\sigma_ys_0$ & $\sigma_xs_0$ & $\sigma_zs_0$ &$i\sigma_0s_y$ &$\sigma_0s_0$ & $\sigma_0s_y$\\
$A_2$ &$\sigma_0s_0$ & $i\sigma_zs_0$ & $\sigma_ys_0$ & $\sigma_xs_0$ & $i\sigma_x s_y$ & 
$\sigma_ys_y$ & $\sigma_zs_0$\\
$B_1$ & $\sigma_0s_0$ & $i\sigma_xs_0$ & $\sigma_zs_0$ & $\sigma_ys_0$ & $i\sigma_zs_y$ & $\sigma_0s_0$ & $\sigma_zs_y$ \\
$B_2$ & $\sigma_0s_0$ & $i\sigma_xs_0$ & $\sigma_ys_0$ & $\sigma_zs_0$ & $i\sigma_zs_y$ & $\sigma_0s_0$ & $\sigma_zs_y$\\ 
\hline
\\
\multicolumn{8}{c}{ $p4g$ $\bar{M}$ point}  \\ \cline{1-8}
 IR of $\Delta$ &
  $U_{\{e|{\bm 0}\},\gamma}$ & 
 $U_{\{4_z|{\bm 0}\},\gamma}$ & 
 $U_{\{2_z|{\bm 0}\},\gamma}$ & 
  $U_{\{\sigma_{(100)}|{\bm \tau}_x+{\bm \tau}_y\},\gamma}$ &
  $T_\gamma$ & $C_\gamma$ & $\Gamma_\gamma$ \\ \hline
 $A_1$ &
 $\sigma_0s_0$&
 $e^{i\pi(2\sigma_0s_z+\sigma_z s_z)/4}$ &
 $-i \sigma_z s_z$ &
$\sigma_x s_0$ &
$i\sigma_0s_y$ & 
$\sigma_0 s_x$ &
$\sigma_0 s_z$\\
$A_2$&
$\sigma_0s_0$&
$e^{i\pi(2\sigma_0s_z+\sigma_z s_z)/4}$ &
$-i \sigma_z s_z$ &
 $\sigma_x s_0$&
 $i\sigma_0 s_y$&
 $\sigma_z s_x$&
 $\sigma_z s_z$ \\
 $B_1$ &
$\sigma_0s_0$&
$e^{i\pi(2\sigma_0s_z+\sigma_z s_z)/4}$ &
$-i \sigma_z s_z$ &
 $\sigma_x s_0$&
 $i\sigma_0 s_y$ &
 $\sigma_x s_0$&
 $\sigma_x s_y$\\ \hline\hline
  \end{tabular}
 \end{center}
 \end{table*}

In the present cases, Eq.(\ref{eq:corr}) is not available. 
To evaluate Eq.(\ref{eq:chiOmega+-}), we explicitly construct the representation $U_{g,\gamma}$,  $T_\gamma$, $C_\gamma$ and $\Gamma_\gamma$ defined below,
\begin{align}
&{\cal U}_g^{\bm k}|u_0^{(a)}\rangle=\sum_b|u_0^{(b)}\rangle[U_{g,\gamma}]_{ba},
\nonumber\\
&{\cal T}|u_0^{(a)}\rangle=\sum_b|u_0^{(b)}\rangle[T_{\gamma}]_{ba},
\nonumber\\
&{\cal C}|u_0^{(a)}\rangle=\sum_b|u_0^{(b)}\rangle[C_{\gamma}]_{ba},
\nonumber\\
&\Gamma|u_0^{(a)}\rangle=\sum_b|u_0^{(b)}\rangle[\Gamma_{\gamma}]_{ba}.
\end{align}
Here $\Gamma_\gamma=-iT_\gamma C_\gamma^*$ since $\Gamma=-i{\cal T}{\cal C}$.
Whereas $U_{g,\gamma}$, $T_\gamma$, $C_\gamma$, and $\Gamma_\gamma$ obey the same multiplication law as ${\cal U}_g^{\bm k}$, ${\cal T}$, ${\cal C}$ and $\Gamma$, the dimension of the representation becomes half, as explained shortly.

 Let us start with a zero mode $|u_0^{(a)}\rangle$ belonging to an irrep $\alpha$ of $G_0^{\bm k}$. 
 Using the Wigner test $W^T_\alpha$,
 we can obtain a co-representation of $G_0^{\bm k}+TG_0^{\bm k}$ from the irrep $\alpha$, in the standard manner.
However,  to obtain a representation of $G^{\bm k}=G_0^{\bm k}+TG_0^{\bm k}+CG_0^{\bm k}+\Gamma G_0^{\bm k}$, we also need to take into account either PHS or CS. 
If either PHS or CS is taken into account, another one is automatically included,  because $\Gamma=-i{\cal T}{\cal C}$.

To include these antisymmetries, we refer to the EAZ class of $\alpha$. 
The zero mode $|u_0^{(a)}\rangle$ can exist in the irrep $\alpha$
when the EAZ class has a non-trivial topological invariant $\mathbb{Z}$ or $\mathbb{Z}_2$.
If  the topological invariant is $\mathbb{Z}$, the irrep $\alpha$ keeps CS, and the zero mode can be an eigenstate of $\Gamma$, say $\Gamma|u_0^{(a)}\rangle=|u_0^{(a)}\rangle$, in a proper basis.
On the other hand, if the topological invariant is $\mathbb{Z}_2$, the irrep $\alpha$ keeps PHS, and the zero mode can satisfy ${\cal C}|u_0^{(a)}\rangle=|u_0^{(a)}\rangle$ in a proper basis.
In this manner, the action of CS or PHS to the zero mode is determined by referring to the EAZ class.
 Using the relation $\{\Gamma, {\cal T}\}=0$ and $[{\cal C}, {\cal T}]=0$,  we can also generalize this argument to the co-representation obtained from $\alpha$, which provides a representation of $G^{\bm k}$.
 
 For instance, let us consider MKPs at the $\bar{M}$ point of $pgg$. 
 We also assume that the irrep of the gap function is $A_1$.
 As a fermion, MFs belong to a double-valued representation, which is uniquely given by the $M$ irrep in Table \ref{tab:nonsym}.
 The $M$ irrep is two-dimensional and obeys
 \begin{align}
&U_{\{2_z|{\bm 0}\},\gamma}^2=-1,  
\nonumber\\
&U_{\{\sigma_{(010)}|{\bm \tau}_x+{\bm \tau}_y\},\gamma}^2=1,  
 \nonumber\\
&U_{\{\sigma_{(100)}|{\bm \tau}_x+{\bm \tau}_y\},\gamma}^2=1,
 \end{align}
 together with the other standard commutation relations between the generators.
 When the gap function belongs to the A$_1$ irrep,  these generators also commute with PHS ${\cal C}$.
 As shown in Table \ref{tab:EAZnonsym}, the Wigner test indicates that the EAZ class is DIII, of which topological invariant is $\mathbb{Z}_2$,  Thus, we can take the basis satisfying the condition ${\cal C}|u_0^{(a)}\rangle=|u_0^{(a)}\rangle$.
 Because $U_{\{\sigma_{(100)}|{\bm \tau}_x+{\bm \tau}_y\},\gamma}$ commutes with ${\cal C}$ and has real eigenvalues $\pm 1$, 
 the eigenbasis of $U_{\{\sigma_{(010)}|{\bm \tau}_x+{\bm \tau}_y\},\gamma}$ satisfies the above condition.
 Moreover, to take into account TRS, we add another $M$ irrep of MFs according to the Wigner test of TRS.
 Consequently, we obtain 
 \begin{align}
&U_{\{2_z|{\bm 0}\},\gamma}=i\sigma_ys_0, 
\ \
U_{\{\sigma_{(010)}|{\bm \tau}_x+{\bm \tau}_y\},\gamma}=\sigma_xs_0,  
\nonumber\\
&U_{\{\sigma_{(100)}|{\bm \tau}_x+{\bm \tau}_y\},\gamma}=\sigma_zs_0,
\ \
C_\gamma=\sigma_0s_0,
\ \ T_\gamma=i\sigma_0s_y.
  \end{align}

 It should be noted here that PHS and CS do not increase the dimension of the representation.
 This is sharp contrast to the representation for ordinary electron systems.
 For ordinary systems, PHS and CS transform an electron to an hole, and thus to realize these symmetries, we need to consider the corresponding holes at the same time, which doubles the degrees of freedom.  
In contrast, in the case of MFs,   we do not need to double the degrees of freedom since MFs are their own anti-particles. These antisymmetries map MFs to themselves.  
As a result, MFs provide a representation shorter than that for ordinary electrons.
The short representation is a central property of MFs, which is a group theoretical manifestation of  the self-conjugate property of MFs.
In a different context, a similar short representation has been known for the BPS states in supersymmetric theories~\cite{Witten78}.
In Table \ref{tab:pgg-p4g-shortrep}, we summarize the short representation of MKPs at the $\bar{M}$ point for $pgg$ and $p4g$.

 \subsection{electric and magnetic responses}
 
 \begin{table*}[t]
\caption{ EAZ symmetry classes, irreps of $\rho^{(ab)}_{\pm}$, and electric and magnetic responses of the double MKPs at the $\bar{M}$ point for $pgg$ and $p4g$.
} \label{tab:elerep}
\begin{center}
  \begin{tabular}{cccccc} \hline\hline
 \multicolumn{6}{c}{  $pgg$ $(C_{2v})$ $\bar{M}$ point, $(M, M)$} \\ \hline
 IR of $\Delta$ 
 & IR of $\rho_+^{(ab)}$ 
 &IR of $\rho_-^{(ab)}$ 
 & Electric multipole $f$ 
 & Electric multipole $g_+$ 
  &Magnetic multipole $g_-$
 \\ \hline
 A$_1$ 
 & 2A$_2$ 
 & A$_1$+A$_2$+B$_1$+B$_2$ 
 & $E_xE_y$ 
 & $B_xB_y$ 
  & $B_x$, $B_y$, $B_z$, $B_xB_yB_z$
  \\
 A$_2$ 
 & B$_1+$B$_2$ 
 & 3A$_1$+A$_2$ 
 & $E_x$, $E_y$ 
 & $B_xB_z$, $B_yB_z$
  & $B_z$, $B_xB_yB_z$
  \\
 B$_1$ 
 & A$_2+$B$_2$ 
 & A$_1$+B$_1$+2B$_2$ 
 & $E_y$, $E_xE_y$ 
 & $B_xB_y$, $B_yB_z$
  & $B_x$, $B_y$, $B_xB_yB_z$
 \\
 B$_2$ 
 & A$_2+$B$_1$ 
 &A$_1$+2B$_1$+B$_2$ 
 & $E_x$, $E_xE_y$ 
 & $B_xB_y$, $B_xB_z$ 
 & $B_x$, $B_y$, $B_xB_yB_z$
 \\ \hline
 \\
 \multicolumn{6}{c}{  $p4g$ $(C_{4v})$ $\bar{M}$ point, $(M_1, M_2)$}  \\ \hline
   IR of $\Delta$ 
   & IR of $\rho_+^{(ab)}$ 
   & IR of $\rho_-^{(ab)}$
   & Electric multipole $f$
   & Electric multipole $g_+$
    & Magnetic multipole $g_-$
    \\ \hline
 A$_1$ 
 & 2B$_2$ 
 &A$_1$+A$_2$+E
 & $E_xE_y$
 & $B_xB_y$ 
  & $B_z$, $\{B_x, B_y\}$, $B_xB_yB_z(B_x^2-B_y^2)$ 
 \\
 A$_2$ 
 & E 
 &A$_1$+A$_2$+2B$_1$
 & $\{E_x,E_y\}$ 
 & $\{B_xB_z,B_yB_z\}$
  & $B_z$, $B_xB_yB_z$, $B_xB_yB_z(B_x^2-B_y^2)$ \\
 B$_1$ 
 & A$_2+$B$_2$ 
 &A$_2$+B$_1$+E
 & $E_xE_y$, $E_xE_y(E_x^2-E_y^2)$ 
 & $B_xB_y$, $B_xB_y(B_x^2-B_y^2)$
  &$B_z$, $\{B_x, B_y\}$, $B_xB_yB_z$\\
 B$_2$ & $-$ & $-$ & $-$ & $-$ & $-$ \\ 
 \hline\hline
 \end{tabular}
\end{center}
\end{table*}

 Using Eq.(\ref{eq:chiOmega+-}), we can determine possible electric and magnetic couplings of the double MKPs at the $\bar{M}$ point for $pgg$ and $p4g$.
 For instance, let us consider $pgg$ and the $A_1$ gap function. 
 From the short representation in Table \ref{tab:pgg-p4g-shortrep}, $\chi_g^{\Omega^{\pm}}$ reads
 \begin{align}
 &
 \chi_{\{e|{\bm 0}\}}^{\Omega^{+}}
 = 2,
  \quad
  \chi_{\{e|{\bm 0}\}}^{\Omega^{-}}
 = 4,
 \nonumber\\
 &\chi_{\{2_z|{\bm 0}\}}^{\Omega^{+}}
 = 2,
  \quad
  \chi_{\{2_z|{\bm 0}\}}^{\Omega^{-}}
 = 0,
  \nonumber\\
  &\chi_{\{\sigma_{(010)}|{\bm \tau}_x+{\bm \tau}_y\}}^{\Omega^{+}}
 = -2,
  \quad
  \chi_{\{\sigma_{(010)}|{\bm \tau}_x+{\bm \tau}_y\}}^{\Omega^{-}}
 = 0,
  \nonumber\\
  &\chi_{\{\sigma_{(100)}|{\bm \tau}_x+{\bm \tau}_y\}}^{\Omega^{+}}
 = -2,
  \quad
  \chi_{\{\sigma_{(100)}|{\bm \tau}_x+{\bm \tau}_y\}}^{\Omega^{-}}
 = 0.
  \end{align}
 Then, using the standard group theoretical method, we perform the irreducible decomposition of $\rho^{(ab)}_\pm$ in terms of the single-valued irreps of $C_{2v}$,
 \begin{align}
 \rho^{(ab)}_+=2\rm A_2,
 \quad
\rho^{(ab)}_-=A_1+A_2+B_1+B_2.
 \end{align}
Because $\rho^{(ab)}_+\neq 0$, in contrast to a single MKP, the double MKPs may host an electric response.
We find that the electric response is quadrupole.
As explained in Sec.\ref{sec:emcoupling},  $f({\bm E})$ in Eq.(\ref{eq:g}) shares the same irrep with 
$\rho^{(ab)}_+$. Thus, referring to irrpes of $f({\bm E})$ in Table \ref{tab:g}, we find
\begin{align}
f^{\rm A_2}_{pgg}({\bm E})=c_2 E_x E_y,
\end{align}
with a constant $c_2$, which is quadrupole.
We can also evaluate the magnetic response of the double MKPs. 
Since $g_+({\bm B})$ and $g_-({\bm B})$ in Eq.(\ref{eq:f+f-}) are possible, their leading terms are given by
\begin{align}
g^{\rm A_2}_{pgg+}({\bm B})&=c_{2+} B_xB_y, \notag \\
g^{\rm A_1}_{pgg-}({\bm B}) &=c_{3-}B_xB_yB_z, \notag \\
g^{\rm A_2}_{pgg-}({\bm B}) &=c_{1,1-}B_z, \notag \\
g^{\rm B_1}_{pgg-}({\bm B})&=c_{1,2-}B_y, \notag \\
g^{\rm B_2}_{pgg-}({\bm B})&= c_{1,3-}B_x,
\end{align}
where $c_{2+}$ $c_{3-}$, and $c_{1,i-}$ are material dependent parameters. This result implies that the leading magnetic response of the double MKPs is a mixture of dipole and quadrupole.

In a similar manner,  we calculate $\chi_g^{\Omega^{\pm}}$ for all possible double MKPs at the $\bar{M}$ point in $pgg$ and $p4g$, and evaluate possible electric and magnetic responses. 
The obtained electric and magnetic responses are summarized in Table.\ref{tab:elerep}.

Before closing this section, we comment on the case of $p4g$, which also realizes a unique electromagnetic response at the $\bar{M}$ point. Although different irreps coexist in the magnetic response, one of them exhibits a magnetic response with high mutipolarity. If the irrep of ${\cal O}_-$ is  A$_1$ or B$_1$, $g_-(\bm{B})$ is of the form: 
\begin{align}
& g_{p4g-}^{\rm A_1} (\bm{B}) = c_{5-} B_x B_y B_z (B_x^2 -B_y^2), 
\\
& g_{p4g-}^{\rm B_1} (\bm{B}) = c_{3-} B_x B_y B_z,
\end{align}
which indicate the magnetic 32-pole and octupole responses, respectively. Similarly, when the irrep of ${\cal O}_+$ is $A_2$, one of electric responses exhibits a 16-pole response,
\begin{align}
&f_{p4g}^{\rm A_2} (\bm{E})= c_4 E_xE_y(E_x^2-E_y^2),
\\
&g_{p4g+}^{\rm A_2} (\bm{B})=c_{4+} B_xB_y(B_x^2-B_y^2).
\end{align}

\section{Summary}
\label{sec:summary}
Applying the Wigner's test to the identification of 1d topological invariants and establishing the multipole theory for MKPs, we classified the possible magnetic structures for MKPs under the wallpaper groups. For a single MKP, irreps of magnetic structures are classified into the magnetic dipole or octupole and one-to-one correspond to those of gap functions in TSCs.  Although almost magnetic structures belong to the magnetic dipole, the magnetic octupole response is realized in  two ways: one is the threefold-rotation-symmetry-induced magnetic octupole in spin $3/2$ TSCs, which is realized for $p6$, $p3m$1, $p31m$ and $p6m$. The magnetic response preserves the sixfold rotation symmetry as shown in Fig.~\ref{fig:p6model} (c) and its shape is described by Eq.~(\ref{eq:MRp3m1}) or (\ref{eq:MRp6}). The other is the glide-symmetry-induced magnetic octupole in nonsymmorphic TSCs, which is realized for $pmg$ and $pgg$ at the BZ boundary. The shape of the magnetic response is given by Eq.~(\ref{eq:MRpmg}) as shown in Fig.~\ref{fig:pmmamodel} (c). In addition, we found that two MKPs arise at the $\bar{M}$ point when the surface BZ preserves $pgg$ or $p4g$. The two MKPs potentially exhibit electric multipole responses, which will be discussed somewhere.

Finally, we comment on the possible experimental method for detecting the magnetic structures of a single MKP.  Our results predict that the spin structure of the MKPs is anisotropic, so we are able to measure the anisotropy through surface-spin-sensitive measurements, such as spin-resolved tunneling spectroscopy~\cite{Jeon17,Cornils17}, spin relaxation rate~\cite{Chung09}, spin susceptibility~\cite{Nagato09}, thermal conductivity~\cite{Nakai14,Xie15,Gnezdilov16} under magnetic fields, and so on. As an example, we discuss the behavior of tunneling conductance under a magnetic field or with a magnet attached, where we assume that only the Zeeman magnetic field affects the MKP. We note that the orbital magnetization is also useful to measure the topological surface states on TSCs~\cite{Tanaka95,Tanaka02,Tanuma02,Tanaka09,Tamura17,Chirolli18}. Tunnel conductance detects the MKP as a zero bias conductance peak~\cite{Tanaka12}. When the magnetic field is turned on,  the MKP shifts from the zero energy, resulting in broadening or splitting of the zero bias conductance peak. When the magnetic structure is the dipole type, such suppression of the zero bias conductance peak can be observed when applying the magnetic field in a specific direction, e.g. a rotation symmetry axis. On the other hand, when the magnetic structure is the octupole type, the suppression occurs in different three directions due to three or sixfold rotation symmetry. Thus, if we apply the in-plane rotating magnetic field, the recovery of the peak may appear along with sixfold periodicity according to Eq.~(\ref{eq:MRp3m1}) or (\ref{eq:MRp6}). 

\acknowledgments 
This work was supported by JSPS KAKENHI (Grants Nos. JP18H04224, JP19K14612, JP20H00131, JP20K03835, and JP20H04635), the Sumitomo Foundation (190228), the CREST project (JPMJCR16F2, JPMJCR19T2) from Japan Science and Technology Agency (JST), and the JSPS Core-to-Core Program (A. Advanced Research Network).

\appendix

\section{Derivation of Eq.~(\ref{eq:gamma-rel})}
\label{app:gamma}
Here we show the derivation of Eq.~(\ref{eq:gamma-rel}). We start with Eq.~(\ref{eq:modeexp}), which can be rewritten as
\begin{align}
 \hat{\gamma}_a &= \int dx \; \langle u_0^{(a)} (x)| \hat{\Psi} (x) \notag \\
&= \int dx \; \sum_{i \tau} (u (x)_0^{(a) \ast})_{i \tau} \hat{\Psi} (x)_{i \tau}, \label{eq:gamma}
\end{align}
where $i$ and $\tau$ describes internal degrees of freedom for electrons and the Nambu space and the wave function satisfies the commutation relation $\left\{ \hat{\Psi}(x)_{i \tau} , \hat{\Psi}^{\dagger}(x')_{j \tau'} \right\} = \delta_{\tau \tau'} \delta_{ij} \delta(x-x')$ and the PHS $\mathcal{C} \hat{\Psi}(x) = \hat{\Psi}(x)$. Similarly, $\hat{\gamma}_a^{\dagger}$ is given by
\begin{align}
 \hat{\gamma}_a^{\dagger} = \int dx \; \langle u_0^{(a) \ast} (x)| \hat{\Psi} (x)^{\dagger}. \label{eq:gammad}
\end{align}
Then, the commutation relation between $\hat{\gamma}_a$ and $\hat{\gamma}_b^{\dagger}$ is calculated as 
\begin{align}
 \left\{\gamma_a , \gamma_b^{\dagger} \right\} &= \int dx dx' \left\{ \langle u_0^{(a)} (x)| \hat{\Psi} (x),  \langle u_0^{(b) \ast} (x')| \hat{\Psi} (x')^{\dagger} \right\} \notag \\
&= \int dx dx' \sum_{ij} \sum_{\tau \tau'} (u (x)_0^{(a) \ast})_{i \tau} (u (x')_0^{(b)})_{j \tau'} \notag \\
& \qquad \qquad \qquad \qquad  \times \left\{ \hat{\Psi} (x)_{i \tau},  \hat{\Psi} (x')^{\dagger}_{j \tau'}  \right\} \notag \\
&= \int dx \;  \sum_{i \tau}  (u (x)_0^{(a) \ast})_{i \tau} (u (x)_0^{(b)})_{i \tau}\notag \\
&= \delta_{ab}  \label{eq:gamma-comm}
\end{align}
Also, from Eq.~(\ref{eq:gammad}), we find the relation between $\gamma_a$ and $\gamma_a^{\dagger}$:
\begin{align}
 \gamma_a^{\dagger} &= \int dx \; \langle u_0^{(a) \ast} (x)| \tau_x \tau_x \hat{\Psi} (x)^{\dagger} \notag \\ 
&= \int dx \; \langle \mathcal{C} u_0^{(a)} (x)|  \hat{\Psi} (x) \notag \\ 
&= \sum_b \gamma_b C_{ba}^{\ast}, \label{eq:gamma-rel1}
\end{align}
where we define a unitary matrix $C_{ba} = \langle u_0^{(b)}| \mathcal{C} u_0^{(a)} \rangle  = C_{ab}$. Using Eqs.~(\ref{eq:gamma-comm}) and (\ref{eq:gamma-rel1}), Eq.~(\ref{eq:gamma-comm}) is recast into
\begin{align}
 \left\{\gamma_a , \gamma_b^{\dagger} \right\} &=\left\{\gamma_a , \gamma_c \right\} C_{cb}^{\ast} \notag \\
&= \delta_{ab},
\end{align}
where $C_{ab}$ satisfies
\begin{align}
C_{ab}^{-1} =C_{ab}^{\dagger} = C_{ba}^{\ast} = C_{ab}^{\ast}, \label{eq:propCab}
\end{align}
so we obtain
\begin{align}
 \left\{\gamma_a , \gamma_b \right\}  = C_{ab}. \label{eq:gamma-comm2}
\end{align}
\section{Derivation of Eq.(\ref{eq:Omega+-})}
\label{app:rep}
In this section, we derive Eq.(\ref{eq:Omega+-}). First, we show the relation
\begin{align}
{\cal T}\rho^{(ab)\dagger}{\cal T}^{-1}=-\Gamma \rho^{(ab)}\Gamma^\dagger,
\quad \Gamma=-i{\cal C}{\cal T}.
\label{eq:rhogamma}
\end{align}
To show this, 
we rewrite the left hand side of the above equation as follows,
\begin{align}
{\cal T}\rho^{(ab)\dagger}{\cal T}^{-1}&=
{\cal U}_T
\left\{
[|u_0^{(a)}\rangle\langle {\cal C}u_0^{(b)}|
-|u_0^{(b)}\rangle\langle{\cal C}u_0^{(a)}|]^{\dagger}
\right\}^*{\cal U}_T^\dagger
\nonumber\\
&={\cal U}_T
\left[|{\cal C}u_0^{(b)}\rangle\langle u_0^{(a)}|
-|{\cal C}u_0^{(a)}\rangle\langle u_0^{(b)}|
\right]^*{\cal U}_T^\dagger
\nonumber\\
&=|{\cal T}{\cal C} u_0^{(b)}\rangle\langle {\cal T}u_0^{(a)}|-
|{\cal T}{\cal C}u_0^{(a)}\rangle\langle{\cal T}u_0^{(b)}|.
\nonumber\\
&=|{\cal T}{\cal C} u_0^{(b)}\rangle\langle {\cal T}{\cal C}{\cal C}u_0^{(a)}|-
|{\cal T}{\cal C}u_0^{(a)}\rangle\langle{\cal T}{\cal C}{\cal C}u_0^{(b)}|.
\end{align}
Then, using ${\cal T}{\cal C}|u_0^{(a)}\rangle=i\Gamma|u_0^{(a)}\rangle$ and ${\cal T}{\cal C}{\cal C}|u_0^{(a)}\rangle=i\Gamma|{\cal C}u_0^{(a)}\rangle$, we obtain Eq. (\ref{eq:rhogamma}).

Since $\Gamma$ anticommutes with the BdG Hamiltonian, if $|u_0^{(a)}\rangle$ is a zero mode, $\Gamma|u_0^{(a)}\rangle$ is also a zero mode.  Thus, it can be written as 
\begin{align}
\Gamma|u_0^{(a)}\rangle=\sum_b |u_0^{(b)}\rangle [\Gamma_\gamma]_{ba}
\end{align}
with $[\Gamma_\gamma]_{ba}=\langle u_0^{(b)}|\Gamma| u_0^{(a)}\rangle$.
From $\{\Gamma, {\cal C}\}=0$, we also have
\begin{align}
\Gamma|{\cal C} u_0^{(a)}\rangle=-\sum_b |{\cal C} u_0^{(b)}\rangle [\Gamma_\gamma]^*_{ba}
\end{align}
Therefore, the right hand side of Eq.(\ref{eq:rhogamma}) is recast into
\begin{align}
-\Gamma\rho^{(ab)}\Gamma^{\dagger}=\sum_{cd}\rho^{(cd)}[\Gamma_\gamma]_{ca}[\Gamma_\gamma]_{db}.
\end{align}
Thus, $\rho_\pm^{(ab)}$ in Eq.(\ref{eq:O+-rho+-}) is rewritten as
\begin{align}
\rho_\pm^{(ab)}&=\frac{1}{2}(\rho^{(ab)}\mp\Gamma\rho^{(ab)}\Gamma^{\dagger})
\nonumber\\
&=\sum_{cd}\rho^{(cd)}P^{\pm}_{(cd)(ab)}
\end{align}
where $P^{\pm}_{(cd)(ab)}$ is defined by
\begin{align}
P^{\pm}_{(cd)(ab)}=\frac{1}{2}(\delta_{ca}\delta_{db}\pm [\Gamma_\gamma]_{ca}[\Gamma_\gamma]_{db}).
\end{align}
Here we find that $P^{\pm}_{(cd)(ab)}$ is a projection; it obeys 
\begin{align}
&P^{+}_{(cd)(ab)} +P^{-}_{(cd)(ab)} =\delta_{ca}\delta_{db},
\nonumber\\
&\sum_{cd}P^{\pm}_{(ef)(cd)}P^{\pm}_{(cd)(ab)}=P^{\pm}_{(ef)(ab)}.
\end{align}
Thus, we also have 
\begin{align}
\rho^{(ab)}_{\pm}=\sum_{cd}\rho^{(cd)}_\pm P^{\pm}_{(cd)(ab)}.
\label{eq:rho_pmP}
\end{align}

Now we derive Eq.(\ref{eq:Omega+-}). 
Using the relation ${\cal U}_g^{\bm k}\Gamma=\eta_g\Gamma {\cal U}_g^{\bm k}$ and Eqs.(\ref{eq:rhogamma}) and (\ref{eq:rep-rhov1}),  we find 
\begin{align}
{\cal U}_g^{\bm k}\rho_{\pm}^{(ab)}{\cal U}_g^{{\bm k}\dagger}
&=
\frac{1}{2}
\left(
{\cal U}_g^{\bm k}\rho^{(ab)}{\cal U}_g^{{\bm k}\dagger}\pm
\Gamma{\cal U}_g^{\bm k}\rho^{(ab)}{\cal U}_g^{{\bm k}\dagger}\Gamma^{\dagger}
\right)
\nonumber\\
&=\sum_{cd}\frac{1}{2}(\rho^{(cd)}\mp\Gamma\rho^{(cd)}\Gamma^{\dagger})[\Omega_g]_{(cd)(ab)}
\nonumber\\
&=\sum_{cd}\rho_{\pm}^{(cd)}[\Omega_g]_{(cd)(ab)}.
\end{align}
From Eq.(\ref{eq:rho_pmP}),  the above equation is recast into
\begin{align}
{\cal U}_g^{\bm k}\rho_{\pm}^{(ab)}{\cal U}_g^{{\bm k}\dagger}
=\sum_{ef}\rho_{\pm}^{(ef)}P^{\pm}_{(ef)(cd)}[\Omega_g]_{(cd)(ab)},
\end{align}
which is nothing but Eq, (\ref{eq:Omega+-}) because it holds that
\begin{align}
\sum_{cd}P^{\pm}_{(ef)(cd)}[\Omega_g]_{(cd)(ab)}=[\Omega^\pm_g]_{(ef)(ab)}.
\end{align}

\section{Double-valued representation of nonsymmorphic wallpaper groups at the BZ boundary}
The double-valued representation of nonsymmorphic wallpaper groups is listed in Table \ref{tab:nonsym}.  

\begin{table}[tbp]
\caption{ Double-valued irreps of nonsymmorphic wallpaper groups at the BZ boundary. In $p4g$, only a minimal set of symmetry operators is shown.
} \label{tab:nonsym}
\begin{center}
  \begin{tabular}{ccccccccc} \hline\hline
   \multicolumn{3}{c}{ $pg$ $\bar{X}$ point} \\ \cline{1-3}
   & $\{e|\bm{0}\}$ & $\{\sigma_{(010)}|\bm{\tau}_x\} $ \\  \cline{1-3}
   $X_1$ & 1 & 1  \\ 
   $X_2$ & 1 & -1 \\ \cline{1-3}
\\
 \multicolumn{5}{c}{ $pmg$  $\bar{X}$ point} \\ \cline{1-5}
             & $\{e|\bm{0}\}$ & $\{2_z| 0\}$ & $\{\sigma_{(010)}|\bm{\tau}_x\}$ & $\{\sigma_{(100)}|\bm{\tau}_x\}$ \\ \cline{1-5}
    $X_1'$ & $1$ & $-i$ & $1$ & $-i$ \\
     $X_2'$ & $1$ & $i$ & $1$ & $i$ \\
     $X_3'$ & $1$ & $i$ & $-1$ & $-i$ \\
     $X_4'$ & $1$ & $-i$ & $-1$ & $i$ \\ \cline{1-5}
 \\
   \multicolumn{5}{c}{ $pgg$ $\bar{M}$ point}  \\ \cline{1-5}
  & $\{e|\bm{0}\}$ & $\{2_z| 0\}$ & $\{\sigma_{(010)}|\bm{\tau}_x +\bm{\tau}_y\}$ & $\{\sigma_{(100)}|\bm{\tau}_x+\bm{\tau}_y\}$  \\ \cline{1-5}
  $M$ & $\sigma_0$ & $-i\sigma_y$ & $\sigma_x$ & $\sigma_z$ \\ \cline{1-5} 
\\
   \multicolumn{5}{c}{ $p4g$ $\bar{M}$ point~\footnote{Note that the basis used here is slightly different from that shown in the Bilbao Crystallographic Server~\cite{Elcoro17}, $P4bm$ (SG\# 100), for a sake of convenience. The two bases are transformed to each other under a unitary transformation.}} \\ \cline{1-5}
  & $\{e|\bm{0}\}$ & $\{4_z| 0\}$ & $\{2_z| 0\}$ & $\{\sigma_{(100)}|\bm{\tau}_x+\bm{\tau}_y\}$\\  \cline{1-5}
  $M_1$ & $\sigma_0$ & $e^{\frac{i\pi}{4}(2\sigma_0 +\sigma_z)}$ & $-i\sigma_z$ & $\sigma_x$ \\
  $M_2$ & $\sigma_0$ & $e^{-\frac{i\pi}{4}(2\sigma_0 +\sigma_z)}$ & $i\sigma_z$ & $\sigma_x$ \\ \hline \hline
 \end{tabular}
\end{center}
\end{table}

\section{Representation of $f_\pm({\bm B})$ and $g({\bm E})$}
Symmetry adopted $f_\pm({\bm B})$ and $g({\bm E})$ are summarized in Tables \ref{tab:magrep}, \ref{tab:g} and \ref{tab:f+}.

\begin{table}[tbp]
\caption{$g_-({\bm B})$ for 2d point groups, and multipole orders. For C$_n$ and C$_{nv}$, we choose the rotation axis as the $z$ axis. For C$_s$, the mirror plane is normal to the $z$ axis.
} \label{tab:magrep}
\begin{center}
  \begin{tabular}{cccc} \hline\hline
$G_0$ & IR & $g_-(\bm{B})$ & multipole order \\ \hline
C$_2$ & A & $B_z$ & Dipole \\
         & B & $B_x$, $B_y$ & Dipole \\
C$_3$ & A & $B_z$ & Dipole \\
         & E & $\{B_x, B_y\}$ & Dipole \\
C$_4$ & A & $B_z$ & Dipole \\
         & B & $B_z (B_x^2-B_y^2)$, $B_y B_y B_z$ & Octupole \\
         & E & \{$B_x$, $B_y$\} & Dipole \\
C$_6$ & A & $B_z$ & Dipole \\
         & B & $B_x^3-3B_x B_y^2$, $B_y^3-3B_y B_x^2$ & Octupole  \\
         & E$_1$ &$\{B_z (B_x^2-B_y^2), B_y B_y B_z\}$ & Octupole  \\
         & E$_2$ & $\{B_x, B_y\}$ & Dipole \\
C$_s$ & A & $B_z$ & Dipole \\
         & B & $B_x$, $B_y$ & Dipole \\
C$_{2v}$ & A$_1$ & $B_x B_y B_z$ & Octupole \\
          & A$_2$ & $B_z$  & Dipole \\
          & B$_1$ & $B_y$  & Dipole \\
          & B$_2$ & $B_x$  & Dipole \\
C$_{3v}$ & A$_1$ &  $B_x^3-3B_x B_y^2$ & Octupole \\
          & A$_2$ & $B_z$  & Dipole \\
          & E & $\{B_x, B_y\}$  & Dipole \\
C$_{4v}$ & A$_1$ & $B_x B_y B_z (B_x^2-B_y^2)$ & 32-pole \\
          & A$_2$ & $B_z$  & Dipole \\
          & B$_1$ & $B_x B_y B_z$  &  Octupole \\
          & B$_2$ & $B_z (B_x^2-B_y^2)$  &  Octupole \\
          & E & $\{B_x, B_y\}$  & Dipole \\
C$_{6v}$ & A$_1$ & $B_z (B_x^3-3B_x B_y^2)(B_y^3-3B_y B_x^2)$ & 128-pole \\
          & A$_2$ & $B_z$  & Dipole \\
          & B$_1$ & $B_y^3-3B_y B_x^2$  &  Octupole \\
          & B$_2$ & $B_x^3-3B_x B_y^2$  &  Octupole \\
          & E$_1$ & $\{B_x, B_y\}$  & Dipole \\
          & E$_2$ & $\{B_z (B_x^2-B_y^2), B_y B_y B_z\}$  & Octupole \\
 \hline\hline
 \end{tabular}
\end{center}
\end{table}

\begin{table}[tbp]
\caption{$f({\bm E})$ for C$_{2v}$ and C$_{4v}$, and multipole orders. We choose the rotation axis as the $z$ axis.}\label{tab:g}
\begin{center}
\begin{tabular}{cccc}\hline\hline
$G_0$ & IR & $f({\bm E})$ & multipole order\\ \hline
C$_{2v}$ & $A_1$ &$E_z$ &Dipole \\
 & $A_2$ &  $E_xE_y$& Quadrupole \\
 & $B_1$ & $E_x$ & Dipole \\
 & $B_2$ & $E_y$ &Dipole \\
 C$_{4v}$ & $A_1$ &$E_z$ & Dipole \\
 & $A_2$ & $E_x E_y(E_x^2-E_y^2)$ & 16-pole \\
& $B_1$ & $E_x^2-E_y^2$ & Quadrupole\\
& $B_2$ & $E_xE_y$ & Quadrupole \\
& $E$ & $(E_x, E_y)$ & Dipole\\ \hline\hline
\end{tabular}
\end{center}
\end{table}

\begin{table}[tbp]
\caption{$g_+({\bm B})$ for C$_{2v}$ and C$_{4v}$, and multipole orders. We choose the rotation axis as the $z$ axis.}\label{tab:f+}
\begin{center}
\begin{tabular}{cccc}\hline\hline
$G_0$ & IR & $g_+({\bm B})$ & multipole order\\ \hline
C$_{2v}$ & $A_1$ &$B_x^2$, $B_y^2$, $B_z^2$ &Quadrupole \\
 & $A_2$ &  $B_xB_y$& Quadrupole \\
 & $B_1$ & $B_xB_z$ & Quadrupole \\
 & $B_2$ & $B_yB_z$ &Quadrupole \\
 C$_{4v}$ & $A_1$ &$B_z^2$, $B_x^2+B_y^2$ & Quadrupole \\
 & $A_2$ & $B_x B_y(B_x^2-B_y^2)$ & 16-pole \\
& $B_1$ & $B_x^2-B_y^2$ & Quadrupole\\
& $B_2$ & $B_xB_y$ & Quadrupole \\
& $E$ & $(B_xB_z, B_yB_z)$ & Quadrupole\\ \hline\hline
\end{tabular}
\end{center}
\end{table}

\section{Topological invariants}
\label{app:top_inv}
We here summarize topological invariants associated with the EAZ class. On a high symmetry point $\bm{k}$ (a high symmetry line $l_{\bm{k}}$), (anti)unitary operators and the BdG Hamiltonian are decomposed into irreps of $G_0^{\bm{k}}$ as Eqs.~(\ref{eq:decompU}) and (\ref{eq:decompBdG}). Then each subsector of the BdG Hamiltonian belongs to the EAZ class and the corresponding topological invariants are defined by using PHS, TRS, and CS projected onto the subsectors. In the following, we define all crystalline symmetry-protected 1d  topological invariants explicitly.

\subsection{$n$-fold rotation symmetry-protected 1d winding number}
Firstly, we define the 1d winding number associated with $n$-fold rotation symmetry ($n =2,3,4,6$). To see this, we assume that the BdG Hamiltonian is invariant under $\mathcal{U}_{\{n_z|\bm{0}\}}$ that satisfies $\mathcal{U}_{\{n_z|\bm{0}\}}^n =-\bm{1}$ and $[\mathcal{U}_{\{n_z|\bm{0}\}},\mathcal{T}]=[\mathcal{U}_{\{n_z|\bm{0}\}},\mathcal{C}]=0$. In this case, the BdG Hamiltonain in the subsectors of $\mathcal{U}_{\{n_z|\bm{0}\}}$ belong to the class AIII; namely, we have an emergent CS $\Gamma^{\alpha}$ within the subsectors. Thus, using $H^{\alpha}(k)$ and $\Gamma^{\alpha}$, the $n$-fold rotation symmetry-protected 1d winding number is defined as~\cite{Mizushima12,Shiozaki14,Dumitrescu14,Lu15,Xiong17}
\begin{align}
 w_{\rm nR}^{\alpha} \equiv \frac{i}{4 \pi} \int^{\pi}_{-\pi} d k \tr \left[ \Gamma^{\alpha} H^{\alpha}(k)^{-1} \partial_k H^{\alpha}(k) \right], \label{eq:wR1D}
\end{align}
where $\alpha$ labes the subsectors of $\mathcal{U}_{\{n_z|\bm{0}\}}$. Equation~(\ref{eq:wR1D}) appears in the wallpaper groups: $p2$, $p3$, $p4$, $p6$, $pmm$, $p31m$, $p3m1$, $p4m$, $p6m$, $pg$, $pmg$, $cmm$ and $p4g$. At the $\bar{M}$ point in the surface BZ, the 4-fold rotation symmetry in $p4g$ leads to $w_{\rm 4R}^{\alpha_0} = -w_{\rm 4R}^{\alpha_1} = w_{\rm 4R}^{\alpha_2} = -w_{\rm 4R}^{\alpha_3}$ for the A$_1$ gap functions and $w_{\rm 4R}^{\alpha_0} = -w_{\rm 4R}^{\alpha_1} = -w_{\rm 4R}^{\alpha_2} = w_{\rm 4R}^{\alpha_3}$ for the $A_2$ gap functions due to TRS and the glide symmetry, where $\alpha_m = \exp \left[ i \pi \left( 2m + 1 \right)/4 \right]$ is an eigenvalue of $\mathcal{U}_{\{4_z|\bm{0}\}}$. Therefore, Majorana zero modes appear as a quartet consisting of two MKPs.
 
\subsection{Mirror-reflection-symmetry-protected 1d winding number} 
 Secondly, we define the 1d winding number associated with mirror-reflection symmetry $\{\sigma|\bm{0}\}$, which is defined in a similar way to the $n$-fold-rotation-symmetry protected 1d winding number. When $\mathcal{U}_{\{\sigma|\bm{0}\}}$ commutes with the BdG Hamiltonian, PHS, and TRS, the EAZ class of BdG Hamiltonain is in class AIII. Hence the mirror-reflection-symmetry-protected 1d winding number is described as
\begin{align}
 w_{\rm \sigma}^{\alpha} \equiv \frac{i}{4 \pi} \int^{\pi}_{-\pi} d k \tr \left[ \Gamma^{\alpha} H^{\alpha}(k)^{-1} \partial_k H^{\alpha}(k) \right], \label{eq:wM1D}
\end{align}
where $\alpha$ is an eigenvalue of $\mathcal{U}_{\{\sigma|\bm{0}\}}$. The 1d winding number appears in the wallpaper groups: $pm$, $pmm$, $p31m$, $p3m1$, $p6m$, $pg$, $cm$, $pmg$, $cmm$, and $pgg$.

\subsection{$n$-fold rotation symmetry-protected 1d $\mathbb{Z}_2$ invariant}
Thirdly, we find $n$-fold rotation symmetry-protected 1d $\mathbb{Z}_2$ invariant in the wallpaper groups: $p2$, $p3$, $p6$, and $p4g$. We define them case by case. For $p2$, the EAZ class of the subsectors of $\mathcal{U}_{\{2_z|\bm{0}\}}$ is D when the irrep of gap functions is B. We have an emergent PHS within the subsectors and a 1d $\mathbb{Z}_2$ invariant associated with the emergent PHS. Using the Berry connection in terms of eigenstates of $H^{\alpha}(k)$, $|u_{n,k}^{\alpha} \rangle$, the 1d $\mathbb{Z}_2$ invariant is defined by
\begin{align}
 \nu_{\rm 2R}^{\alpha} \equiv \frac{1}{\pi} \int^{\pi}_{-\pi} dk \mathcal{A}^{\alpha}(k) \mod 2, \label{eq:nuM1D}
\end{align}
with
\begin{align}
 \mathcal{A}^{\alpha} (k)=-i \sum_{n \in {\rm occ}} \; \langle u_{n, k}^{\alpha} | \partial_k | u_{n,k}^{\alpha} \rangle,
\end{align}
where $\alpha$ is an eigenvalue of $\mathcal{U}_{\{2_z|\bm{0}\}}$ and the summation is taken over the occupied state with a fixed $\alpha$. $ \nu_{\rm 2R}^{\alpha} =  -\nu_{\rm 2R}^{-\alpha}$ is satisfied due to TRS. 

For $p3$ and $p6$, a 1d $\mathbb{Z}_2$ invariant exists only when the spin of electrons is $3/2$. The 3-fold rotation operator $\mathcal{U}_{\{3_z|\bm{0}\}}$ has a real eigenvalue and always commutes with $\mathcal{T}$ and $\mathcal{C}$, so that the subsector of $\mathcal{U}_{\{3_z|\bm{0}\}}$ belongs to class DIII. The 1d $\mathbb{Z}_2$ invariant $\nu_{\rm 3R}^{\alpha}$ is defined by
\begin{align}
 \nu_{\rm 3R}^{\alpha} = \frac{1}{2\pi} \int^{\pi}_{-\pi} dk \mathcal{A}^{\alpha}(k) \mod 2, \label{eq:nu3R}
\end{align}
where the basis of $\mathcal{A}^{\alpha}$ is an eigenstate of $\mathcal{U}_{\{3_z|\bm{0}\}}$ and the gauge fixing condition $\mathcal{T}|u_{2n-1,k}^{\alpha}\rangle = |u_{2n,-k}^{\alpha}\rangle $ is imposed. Similarly, the 6-fold rotation operator $\mathcal{U}_{\{6_z|\bm{0}\}}$ satisfies $\mathcal{U}_{\{6_z|\bm{0}\}}^2 = -\bm{1}$ and $\{\mathcal{U}_{\{6_z|\bm{0}\}}, \mathcal{C}\}=0$, when the spin of electrons is $3/2$ and the irrep of gap function is B. Thus, the subsectors of $\mathcal{U}_{\{6_z|\bm{0}\}}$ belongs to class D, so the 1d $\mathbb{Z}_2$ invariant $\nu_{\rm 6R}^{\alpha}$ is defined by Eq.~(\ref{eq:nuM1D}), where $\alpha$ is an eigenvalue of $\mathcal{U}_{\{6_z|\bm{0}\}}$. 

On the other hand, for p4g, a glide symmetry plays an important role. A 1d $\mathbb{Z}_2$ invariant is defined at the $\bar{M}$ point of $p4g$ when the irrep of the gap function is B$_1$. Here $p4g$ consists of the 4-fold rotation operator $\{4_z|\bm{0}\}$ and the glide operator $\{\sigma_{(010)}|\bm{\tau}_x+\bm{\tau}_y\}$. For the B$_1$ gap function, the PH operator satisfies $\{\mathcal{U}_{\{4_z|\bm{0}\}},\mathcal{C}\}=[\mathcal{U}_{\{\sigma_{(010)}|\bm{\tau}_x+\bm{\tau}_y\}}, \mathcal{C}]=0$. Thus, there is an emergent PHS operator $\mathcal{C}' = \mathcal{C}\mathcal{U}_{\{\sigma_{(010)}|\bm{\tau}_x+\bm{\tau}_y\}}$ with $(\mathcal{C}')^2=1$ for each subsector of $\{4_z|\bm{0}\}$, resulting in that the EAZ class is D.
Therefore, the 1d $\mathbb{Z}_2$ invariant $\nu_{4R}^{\alpha}$ is defined in a similar manner to Eq.~(\ref{eq:nuM1D}), 
where the basis of $\mathcal{A}^{\alpha}$  is an eigenstate of $\mathcal{U}_{\{4_z|\bm{0}\}}$. Moreover, TRS and the other SG operators impose additional constraints on $\nu_{4R}^{\alpha}$, which leads to $\nu_{\rm 4R}^{\alpha_0} =\nu_{\rm 4R}^{\alpha_1} = -\nu_{\rm 4R}^{\alpha_2} = -\nu_{\rm 4R}^{\alpha_3}$, i.e., two MKPs appears.

\subsection{Mirror-reflection symmetry-protected 1d $\mathbb{Z}_2$ invariant} 
 Fourthly, we find a 1d $\mathbb{Z}_2$ invariant associated with mirror-reflection symmetry $\{\sigma|\bm{0}\}$, which is defined in a similar way to the $n$-fold rotation symmetry-protected 1d $\mathbb{Z}_2$ invariant. When $\mathcal{U}_{\{\sigma|\bm{0}\}}$ anticommutes with PHS, the EAZ class of the subesectors becomes class D. That is to say, the mirror-reflection symmetry-protected 1d $\mathbb{Z}_2$ invariant $\nu_{\sigma}^{\alpha}$ is given by Eq.~(\ref{eq:nuM1D}) in terms of $\mathcal{A}^{\alpha}(k)$ in the subsectors of $\mathcal{U}_{\{\sigma|\bm{0}\}}$. The 1d $\mathbb{Z}_2$ appears in the wallpaper groups: $pm$, $p31m$, $p3m1$, $pg$, $cm$, and $pmg$.

\subsection{Glide symmetry-protected 1d $\mathbb{Z}_2$ invariant}
Finally, we define the glide symmetry-protected 1d $\mathbb{Z}_2$ invariant~\cite{QZWang16,Shiozaki16}. To show this, we consider $pg=\{ \{e|\bm{0}\}, \{\sigma_{(010)}|\bm{\tau}_x\} \}$, where $\sigma_{(010)}$ is the mirror reflection with respect to the $(010)$ plane and $\bm{\tau}_x$ is a half translation in the $x$ direction. At the $\bar{X}$ point or the $\bar{M}$ point in the surface BZ, a nontrivial factor system arises as $z_{\{\sigma_{(010)}|\bm{\tau}_x\},\{\sigma_{(010)}|\bm{\tau}_x\}} =1$, so the glide operator satisfies $\mathcal{U}_{\{\sigma_{(010)}|\bm{\tau}_x\}}^2=\bm{1}$ and its eigenvalue is real.  When the BdG Hamiltonian is invariant under $\mathcal{U}_{\{\sigma_{(010)}|\bm{\tau}_x\}}$ and the irrep of the gap function is A, we can define the following  1d topological invariant at $k_x=\pi$: As $[\mathcal{U}_{\{\sigma_{(010)}|\bm{\tau}_x\}}, \mathcal{C}] = [\mathcal{U}_{\{\sigma_{(010)}|\bm{\tau}_x\}}, \mathcal{T}] =0$, PHS and TRS are retained in the subsector of $\mathcal{U}_{\{\sigma_{(010)}|\bm{\tau}_x\}}$ and thus the EAZ class is DIII. The glide symmetry-protected 1d $\mathbb{Z}_2$ invariant is defined as
\begin{align}
 \nu_{\rm G}^{\alpha} = \frac{1}{2\pi} \int^{\pi}_{-\pi} dk \mathcal{A}^{\alpha}(k) \mod 2, \label{eq:nuG1D}
\end{align}
where $|u_{n,k}^{\alpha} \rangle_{\pm}$ is an eigenstate of $\mathcal{U}_{\{\sigma_y|\bm{\tau}_x\}}$ and the gauge fixing condition $\mathcal{T}|u_{2n-1,k}^{\alpha}\rangle_{\pm} = |u_{2n,-k}^{\alpha} \rangle_{\pm} $ is imposed. The $\mathbb{Z}_2$ invariant in Eq.~(\ref{eq:nuG1D}) appears in the wallpaper groups: $pg$ and $pgg$. For the $\bar{M}$ point of $pgg$, we have an additional SG operator anticommuting with the glide operator, which yields $\nu_{\rm G1D}^{\alpha} = \nu_{\rm G1D}^{-\alpha} $ and thus two MKPs appear there.

\section{Enhancement of rotational symmetry}
 \label{app:enhance}
We show here the enhancement of rotational symmetry. 
We start with a time-reversal invariant effective Hamiltonian for spin $j_z$ electrons.
Because of TRS,  the Hamiltonian minimally consists of spin $\pm j_z$ electrons, and thus it is 
given by a $2\times 2$ matrix 
\begin{align}
H(\bm{k}) &= a_0(\bm{k}) s_0 + a_x(\bm{k}) s_x  + a_y(\bm{k}) s_y +  a_z(\bm{k}) s_z \notag \\
             &= a_0(\bm{k}) s_0 + a_-(\bm{k}) s_+  + a_+(\bm{k}) s_- +  a_z(\bm{k}) s_z,
             \label{eq:generalHami}
\end{align}
where $s_i$ are the Pauli matrices acting on the spin space $(|j_z\rangle, |-j_z\rangle)$, $a_i (\bm{k})$ are real functions of $\bm{k}$, and $a_{\pm} = a_x \pm i a_y$ and $s_{\pm} = (s_x \pm i s_y)/2$.   

Then, let us assume that the Hamiltonian in Eq.~(\ref{eq:generalHami}) is invariant under 
the $n$-fold rotation $\{n_z|{\bm 0}\}$ with respect to the $z$ axis:
\begin{align}
&U_{\{n_z|\bm{0}\}}H(\bm{k}) U_{\{n_z|\bm{0}\}}^{\dagger} = H(\{n_z|\bm{0}\}  \bm{k}), \label{eq:C3vHami}
\end{align} 
where $\{n_z|\bm{0}\}{\bm k}=( e^{i 2\pi/n}k_+, e^{-i 2\pi/n} k_- , k_z)$ and $U_{\{n_z|\bm{0}\}}$ is of the form:
\begin{align}
 &U_{\{n_z|\bm{0}\}} = \diag (e^{-i \frac{2\pi}{n}j_z },e^{i \frac{2\pi}{n} j_z}). \label{eq:C3vop}
\end{align}
In order for Eq.~(\ref{eq:generalHami}) to satisfy Eq.~(\ref{eq:C3vHami}), each coefficient in Eq.~(\ref{eq:generalHami}) should satisfy
\begin{subequations}
\begin{align}
 &a_0(\bm{k}) = a_0(\{n_z|\bm{0}\}  \bm{k}), \\
 &e^{-i \frac{4\pi}{n}j_z } a_-(\bm{k}) = a_-(\{n_z|\bm{0}\}  \bm{k}), \\
 &e^{i \frac{4\pi}{n}j_z } a_+(\bm{k}) = a_+(\{n_z|\bm{0}\}  \bm{k}), \\
  &a_z(\bm{k}) = a_z(\{n_z|\bm{0}\}  \bm{k}).
\end{align}
\end{subequations}
Thus, if $e^{i 4\pi j_z/n } \neq 1$, 
$a_+$ and $a_-$ vanish at the high symmetry line $(0,0,k_z)$.
This implies that at the high symmetry line,  
the $n$-fold rotation symmetry becomes the continuous rotation one, and the Hamiltonian is invariant under any rotation with respect to the $z$-direction, 
 \begin{align}
   U_{\theta} H(0,0,k_z) U_{\theta}^{\dagger} = H(0,0,k_z), 
\end{align} 
where $U_{\theta} = \diag (e^{-i \theta j_z },e^{i \theta j_z})$ ($0 \le \theta < 2 \pi$). 
For $j_z=1/2$ and $j_z=5/2$, the condition $e^{i 4\pi j_z/n}\neq1$ is met for any $n=2,3,4,6$, and thus the enhancement of rotation symmetry in the above always occurs.
On the other hand, for $j_z=3/2$, the condition is met only for $n=2,4,6$. 
The enhancement of rotation symmetry does not occur for $j_z=3/2$ with $n=3$.

The enhancement of rotation symmetry may provide an additional protection for MKPs.
For instance, let us consider a topological superconductor hosting a MKP on a surface with 3-fold rotation symmetry.
In the presence of a finite magnetic field parallel to the surface, the 3-fold rotation symmetry is explicitly broken, 
but if the  system consists of $j_z=1/2$ or $j_z=5/2$ electrons, we may retain an additional symmetry: 
By combining 2-fold rotation symmetry obtained by the symmetry enhancement with TRS,  the system supports magnetic 2-fold rotation symmetry.  The magnetic rotation symmetry may stabilize the surface MKP \cite{Mizushima12}.
Such stabilization is expected for a MKP on the $(111)$ surface of superconducting topological insulator 
Cu$_x$Bi$_2$Se$_3$ with the $A_{1u}$ gap function \cite{Fu10}.

\section{Superconducting nodes and Majorana multipole response}
\label{app:node}
In our theory, we implicitly assume that there is no node on the high symmetry line $l_{\bm k}$ in the bulk BZ where the 1d topological invariant is defined. In the following, we show the topological classification for superconducting nodes, which is performed in a similar manner to the topological classification of 1d topological invariants, and discuss when this assumption is satisfied. 
In topological arguments, a stable node on $l_{\bm k}$ is classified by a 0d topological invariant.
The possible 0d topological invariant is specified by the Wigner's test in Eqs.~(\ref{eq:WT}), (\ref{eq:WC}), and (\ref{eq:WG}), where only symmetries that keep a position of the node are taken into account.  

In the following, we classify possible nodes on the high symmetry line for systems (i) without and (ii) with spatial inversion symmetry, respectively. The results are summarized in Table~\ref{tab:nodes}, where possible 0d topological invariants under the wallpaper groups are classified in superconductors without spatial inversion symmetry, even-parity superconductors, and odd-parity superconductors.

{\it (i) systems without inversion symmetry.} --  In addition to $G_0^{\bm k}$, 
we need to consider CS, both of which keep any point on $l_{\bm k}$ invariant. 
The total group we consider is 
\begin{align}
 G^{\bm k}_0 + \Gamma G^{\bm k}_0,
\end{align}
which implies that the Wigner's test is determined solely by $W^\Gamma_\alpha$ in Eq.~(\ref{eq:WG}). 
By forgetting $W^T_{\alpha}$ and $W^C_{\alpha}$, 
the emergent AIII, BDI, DIII, CI, and CII classes in Tables~\ref{tab:EAZsym} and~\ref{tab:EAZnonsym} 
change to  the emergent AIII class.
In these cases, no 0d  topological invariant exists, and thus the system on $l_{\bm k}$ is fully gapped in general.
On the other hand,  the emergent A, AI, AII, D, and C classes in Tables~\ref{tab:EAZsym} and~\ref{tab:EAZnonsym} 
change to the A class. As the A class  hosts a 0d topological invariant, the latter emergent classes may have a stable node on $l_{\bm k}$.
Among these emergent classes, only the D class has a 1d topological invariant at the same time. 

\begin{table*}[tbp]
\caption{ Classification of point nodes under the wallpaper groups. For each table, the first, second, third, and forth columns show irreps of gap functions, emergent Altland Zirnbauer classes for systems without spatial inversion symmetry, with even-parity pairings, and with odd-parity pairings, respectively. Here numbers in parentheses represent 0d topological invariants.    
} \label{tab:nodes}
\begin{center}
  \begin{tabular}{ccccp{2mm}ccccp{2mm}cccc} \hline\hline
   \multicolumn{4}{c}{ $p1$ (C$_1$), spin 1/2}
   && \multicolumn{4}{c}{ $p2$ (C$_2$), spin 1/2}  
   &&  \multicolumn{4}{c}{ $p3$ (C$_3$), spin 1/2} 
   \\ \cline{1-4}\cline{6-9}\cline{11-14}
   IR of $\Delta$ & w/o IS & even parity  & odd parity  
   && IR of $\Delta$ & w/o IS & even parity  & odd parity  
   &&  IR of $\Delta$ & w/o IS & even parity  & odd parity  
   \\  \cline{1-4}\cline{6-9}\cline{11-14}
   A  & AIII($0$) & DIII($0$) & CII($0$) 
   &&  A & AIII($0$) & AIII($0$)  & AIII($0$)    
   &&  A  & AIII($0$) & AIII($0$)  & AIII($0$) 
  \\
   & & &     
  &&   B  & A($\mathbb{Z}$)   & D($\mathbb{Z}_2$) & C($0$)   
  &&       &       &                                          &     
  \\
  \\
   \multicolumn{4}{c}{ $p3$ (C$_3$), spin 3/2}
   &&  \multicolumn{4}{c}{ $p4$ (C$_4$), spin 1/2 or 3/2} 
   &&  \multicolumn{4}{c}{ $p6$ (C$_6$), spin 1/2 or 5/2}  
   \\ \cline{1-4}\cline{6-9}\cline{11-14}
   IR of$\Delta$ &  w/o IS & even parity  & odd parity    
   && IR of $\Delta$ &  w/o IS & even parity  & odd parity 
   && IR of $\Delta$ &  w/o IS & even parity  & odd parity 
   \\  \cline{1-4}\cline{6-9}\cline{11-14}
   A   & AIII($0$)  & DIII($0$)  &  CII($0$) 
   &&  A  & AIII($0$)  & AIII($0$)   & AIII($0$)   
   &&  A  & AIII($0$)   & AIII($0$)   &  AIII($0$)     
   \\
    &  &   &       
   &&  B  & A($\mathbb{Z}$)    & A($\mathbb{Z}$)       &A($\mathbb{Z}$)    
   &&  B   & A($\mathbb{Z}$)     & A($\mathbb{Z}$)    & A($\mathbb{Z}$)      
   \\
    \\
   \multicolumn{4}{c}{$p6$ (C$_6$), spin 3/2} 
   && 
   \multicolumn{4}{c}{ $pm$ (C$_{s}$), spin 1/2} 
   && 
   \multicolumn{4}{c}{ $pmm$ (C$_{2v}$), spin 1/2} 
   \\  \cline{1-4}\cline{6-9}\cline{11-14}
   IR of $\Delta$ &w/o IS & even parity  & odd parity 
   && IR of$\Delta$ & w/o IS & even parity  & odd parity 
   && IR of $\Delta$ &w/o IS & even parity  & odd parity 
   \\  \cline{1-4}\cline{6-9}\cline{11-14}
   A & AIII($0$)   &  AIII($0$)      &AIII($0$)     
   &&   A  & AIII($0$)  & AIII($0$)       & AIII($0$)    
   && A$_1$  & AIII($0$)   & CI($0$)     &BDI($\mathbb{Z}_2$)     
   \\
   B  & A($\mathbb{Z}$)   & D($\mathbb{Z}_2$) & C($0$)      
   &&   B  & A($\mathbb{Z}$)   & D($\mathbb{Z}_2$) & C($0$)   
   &&  A$_2$    & AIII($0$)  & BDI($\mathbb{Z}_2$)   & CI($0$)    
   \\
        &      &                                                 &           
   &&        &      &                                               &        
   &&  B$_1$    & AIII($0$)  & BDI($\mathbb{Z}_2$)   & CI($0$)   
   \\
        &      &                                                 &            
   &&        &      &                                              &        
   &&  B$_2$    & AIII($0$)  & BDI($\mathbb{Z}_2$)   & CI($0$)   
   \\
   \\
  \multicolumn{4}{c}{ $p31m$, $p3m1$  (C$_{3v}$), spin 1/2}  
  && 
  \multicolumn{4}{c}{ $p31m$, $p3m1$ (C$_{3v}$), spin 3/2} 
  && \multicolumn{4}{c}{ $p4m$ (C$_{4v}$), spin 1/2 or 3/2} 
  \\ \cline{1-4}\cline{6-9}\cline{11-14}
   IR of $\Delta$ & w/o IS & even parity  & odd parity 
   && IR of $\Delta$ & w/o IS & even parity  & odd parity 
   && IR of $\Delta$ & w/o IS & even parity  & odd parity 
   \\  \cline{1-4}\cline{6-9}\cline{11-14}
  A$_{1}$ & AIII($0$)   & CI($0$)     &BDI($\mathbb{Z}_2$)  
   &&  A$_{1}$ &AIII($0$)   & AIII($0$)   &  AIII($0$)     
   &&  A$_{1}$ & AIII($0$)   & CI($0$)     &BDI($\mathbb{Z}_2$)  
   \\
  A$_{2}$ &AIII($0$)  & BDI($\mathbb{Z}_2$)   & CI($0$)  
  && A$_{2}$ & A($\mathbb{Z}$)   & D($\mathbb{Z}_2$) & C($0$)    
  &&  A$_{2}$ & AIII($0$)  & BDI($\mathbb{Z}_2$)   & CI($0$)  
  \\
            &      &   &         
  &&       &     &       &         
  &&  B$_{1}$ & A($\mathbb{Z}$)  & AI($\mathbb{Z}$)   &AI($\mathbb{Z}$)    
  \\
            &      &       &       
  &&       &     &       &         
  &&  B$_{2}$ & A($\mathbb{Z}$)  & AI($\mathbb{Z}$)   &AI($\mathbb{Z}$)  
  \\
   \\
 \multicolumn{4}{c}{ $p6m$  (C$_{6v}$), spin 1/2 or 5/2} 
 && \multicolumn{4}{c}{ $p6m$ (C$_{6v}$), spin 3/2}  
 && \multicolumn{4}{c}{ $pg$ (C$_s$) $\bar{X}$ point}
 \\ \cline{1-4}\cline{6-9}\cline{11-14}
   IR of $\Delta$ & w/o IS & even parity  & odd parity        
    && IR of $\Delta$ & w/o IS & even parity  & odd parity 
   && IR of $\Delta$ & w/o IS & even parity  & odd parity  
  \\  \cline{1-4}\cline{6-9}\cline{11-14}
   A$_{1}$ & AIII($0$)   & CI($0$)     &BDI($\mathbb{Z}_2$)   
   && A$_{1}$ & AIII($0$)   & CI($0$)     &BDI($\mathbb{Z}_2$)   
   && A & AIII($0$)   &AIII($0$)    &AIII($0$)
    \\
   A$_{2}$ &AIII($0$)  & BDI($\mathbb{Z}_2$)   & CI($0$)  
   & & A$_{2}$ & AIII($0$)  & BDI($\mathbb{Z}_2$)   & CI($0$) 
   & & B & A($\mathbb{Z}$)  & D($\mathbb{Z}_2$)   & C($0$) 
    \\
   B$_{1}$ & A($\mathbb{Z}$)  & AI($\mathbb{Z}$)   &AI($\mathbb{Z}$)    
    & & B$_{1}$ &AIII($0$)  & BDI($\mathbb{Z}_2$)   & CI($0$) 
    &&       &     &       &     
     \\
   B$_{2}$ &  A($\mathbb{Z}$)  & AI($\mathbb{Z}$)   &AI($\mathbb{Z}$)    
    & & B$_{2}$ &AIII($0$)  & BDI($\mathbb{Z}_2$)   & CI($0$)  
    &&       &     &       &     
     \\  
        \\
   \multicolumn{4}{c}{ $pmg$ (C$_{2v}$) $\bar{X}$ point} 
   && \multicolumn{4}{c}{ $pgg$ (C$_{2v}$) $\bar{M}$ point}  
   && \multicolumn{4}{c}{ $p4g$ (C$_{4v}$) $\bar{M}$ point}
   \\ \cline{1-4}\cline{6-9}\cline{11-14}
   IR of $\Delta$ & w/o IS & even parity  & odd parity        
    && IR of $\Delta$ & w/o IS & even parity  & odd parity 
   && IR of $\Delta$ & w/o IS & even parity  & odd parity  
   \\  \cline{1-4}\cline{6-9}\cline{11-14}
   A$_{1}$ & AIII($0$)   &AIII($0$)    &AIII($0$) 
   && A$_{1}$ & AIII($0$)  & DIII($0$)  &  CII($0$) 
   && A$_{1}$ & AIII($0$)   &AIII($0$)    &AIII($0$)
    \\
   A$_{2}$ &A($\mathbb{Z}$)     & A($\mathbb{Z}$)    & A($\mathbb{Z}$)  
   & & A$_{2}$ & AIII($0$)  & CII($0$)  &  DIII($0$) 
   & & A$_{2}$ & AIII($0$)   &AIII($0$)    &AIII($0$) 
    \\
   B$_{1}$ & A($\mathbb{Z}$)     & A($\mathbb{Z}$)    & A($\mathbb{Z}$)     
    & & B$_{1}$ &AIII($0$)  & DIII($0$)  &  CII($0$) 
    &&  B$_{1}$ &  A($\mathbb{Z}$)     & D($\mathbb{Z}_2$)    & C($0$) 
     \\
   B$_{2}$ &  A($\mathbb{Z}$)     & D($\mathbb{Z}_2$)    & C($0$)   
    & & B$_{2}$ &AIII($0$)  & DIII($0$)  &  CII($0$) 
    &&  B$_{2}$ & A($\mathbb{Z}$)  & C($0$)  &  D($\mathbb{Z}_2$) 
     \\            
 \hline\hline
 \end{tabular}
\end{center}
\end{table*}

{\it (ii) systems with inversion symmetry.} --  
Next, we take into account spatial inversion $\{I|\bm{0}\}$.  Combining TRS and PHS with space inversion, we have $\mathfrak{C} \equiv \{I|\bm{0}\}C$ and $\mathfrak{T} \equiv \{I|\bm{0}\}T$, respectively, both of which keep any point on $l_{\bm k}$ invariant.
The total group relevant to the node stability is
\begin{align}
G^{\bm k}_0 + \mathfrak{T} G^{\bm k}_0 + \mathfrak{C} G^{\bm k}_0 + \Gamma G^{\bm k}_0,  \label{eq:allGnode}
\end{align}
and the Wignar's test for $\mathfrak{T}$ and $\mathfrak{C}$ is given by~\cite{Sumita19}
\begin{align}
 &W_{\alpha}^{\mathfrak{T}} \equiv \frac{1}{|G_0|} \sum_{g \in G_0} z_{\mathfrak{T}g,\mathfrak{T}g} \chi[U^{\alpha}_{(\mathfrak{T}g)^2}]=\pm1,0, \label{eq:WTI} \\
 &W_{\alpha}^{\mathfrak{C}} \equiv \frac{1}{|G_0|} \sum_{g \in G_0} z_{\mathfrak{C}g,\mathfrak{C}g} \chi[U^{\alpha}_{(\mathfrak{C}g)^2}]=\pm1,0, \label{eq:WCI} 
\end{align}
where $\mathfrak{T}^2=z_{\mathfrak{T},\mathfrak{T}}=-1$, $\mathfrak{C}^2=z_{\mathfrak{C},\mathfrak{C}} = \eta_I$ and $\eta_I=1$ ($-1$) indicates an even (odd) parity gap function.
First, we apply the Wigner's test in Eqs.~(\ref{eq:WTI}), (\ref{eq:WCI}), and (\ref{eq:WG}) to symmorphic wallpaper groups. In these cases, the Wigner's test reads
\begin{align}
 (W_{\alpha}^{\mathfrak{T}},W_{\alpha}^{\mathfrak{C}},W_{\alpha}^{\Gamma}) =  (W_{\alpha}^{T}, \eta_I W_{\alpha}^{C},W_{\alpha}^{\Gamma}),
\label{eq:Wtinv}
\end{align}
which determines the EAZ classes for the nodal structure on $l_{\bm k}$.
Note that the EAZ classes are different from those in Tables \ref{tab:EAZsym} and \ref{tab:nonsym} only for the odd-parity superconductors.

From this result, we find that the 1d topological invariants on $l_{\bm k}$ is generally well-defined for odd parity superconductors:  
The 1d topological invariants can be nonzero when the EAZ in Table \ref{tab:EAZsym} and \ref{tab:EAZnonsym} is AIII, BDI, D, DIII, or CII classes, and from Eq.(\ref{eq:Wtinv}), these classes correspond to AIII, CI, C, CII, and DIII, respectively.
Because the latter EAZ classes do not have 0d topological invariants, no stable node appear on $l_{\bm k}$.
On the other hand, for even parity superconductors, the emergent BDI class has both 1d and 0d topological invariants. 
In this case, we need to avoid stable nodes to define the 1d topological invariant.

For nonsymmorphic groups, we need to perform the Wigner's test case by case. First, we consider $pg$ at the $\bar{X}$ point, which is given by $\{ \{e|0\}, \{ \sigma_{(010)}|\bm{\tau}_x \} \}$. 
We obtain
\begin{align}
  (W_{\alpha}^{\mathfrak{T}},W_{\alpha}^{\mathfrak{C}},W_{\alpha}^{\Gamma}) = \left(0,\frac{\eta_I}{2} (1-\eta_{\sigma_{(010)}}), \frac{1}{2} (1+\eta_{\sigma_{(010)}}) \right).
\end{align}
The system has a nontrivial 1d topological invariant for the $A$ gap function ($\eta_{\sigma_{(010)}}=1$). (See Table \ref{tab:nonsym}.) The corresponding EAZ for the node structure is AIII, irrespective of $\eta_I$. Thus, no point node appears. 

For $pmg =\{\{e|0\}, \{2_z|0\},\{\sigma_{(010)}|\bm{\tau}_x\}, \{\sigma_{(100)}|\bm{\tau}_x\} \}$ at the $\bar{X}$ point, the Wigner's test becomes
\begin{widetext}
\begin{align}
  (&W_{\alpha}^{\mathfrak{T}},W_{\alpha}^{\mathfrak{C}},W_{\alpha}^{\Gamma})= \left(0,\frac{\eta_I}{2} (1-\eta_{2_z}-\eta_{\sigma_{(010)}}+\eta_{\sigma_{(100)}}), \frac{1}{4} (1+\eta_{2_z}+\eta_{\sigma_{(010)}}+\eta_{\sigma_{(100)}})\right).
\end{align}
\end{widetext}
In this case, the system has a nontrivial 1d topological invariant for the $A_1$ gap function ($\eta_{2_z}=\eta_{\sigma_{(010)}}=\eta_{\sigma_{(100)}}=1$) or the B$_1$ gap function ($-\eta_{2_z}=\eta_{\sigma_{(010)}}=-\eta_{\sigma_{(100)}}=1$). The EAZ class for the node structure is AIII (A) for the A$_1$ (B$_1$) gap function, regardless of $\eta_I$, and thus a stable node appears for the B$_1$ gap function, which should be avoided to define the 1d topological invariant on $l_{\bm k}$.
A similar node appears for $pgg$ and $p4g$ at the $\bar{X}$ point.

For $pgg =\{\{e|0\}, \{2_z|0\}, \{\sigma_{(010)}|\bm{\tau}_x+\bm{\tau}_y\}, \{\sigma_{(100)}|\bm{\tau}_x+\bm{\tau}_y\} \}$ at the $\bar{M}$ point, we have 
\begin{align}
  (W_{\alpha}^{\mathfrak{T}},W_{\alpha}^{\mathfrak{C}},W_{\alpha}^{\Gamma}) = \left(-1,\frac{\eta_I}{2} (1-\eta_{2_z}+\eta_{\sigma_{(010)}}+\eta_{\sigma_{(100)}}), 1\right).
\end{align}
For any gap function, the EAZ class for the node structure is DIII or CII, and thus no 0d topological invariant exists.

Finally, we consider $p4g$ at the $\bar{M}$ point, which is generated by $\{\{2_z|0\}, \{4_z^+|0\}, \{\sigma_{(010)}|\bm{\tau}_x+\bm{\tau}_y\}, \{\sigma_{(110)}|\bm{\tau}_x+\bm{\tau}_y\}\}$. The Wigner's test is
\begin{align}
  (&W_{\alpha}^{\mathfrak{T}},W_{\alpha}^{\mathfrak{C}},W_{\alpha}^{\Gamma}) \notag \\
 &= \left(0,\frac{\eta_I}{4} (1-\eta_{2_z}+2\eta_{\sigma_{(010)}}-2\eta_{\sigma_{(110)}}), \frac{1}{8} (4+ 4 \eta_{4_z^+})\right).
\end{align}
We have a nontrivial 1d topological invariant for the A$_1$ ($\eta_{4_z^+}=\eta_{2_z}=\eta_{\sigma_{(010)}}=\eta_{\sigma_{(110)}}=1$), the A$_2$ ($\eta_{4_z^+}=\eta_{2_z}=-\eta_{\sigma_{(010)}}=-\eta_{\sigma_{(110)}}=1$), and the $B_1$ gap functions ($-\eta_{4_z^+}=\eta_{2_z}=\eta_{\sigma_{(010)}}=-\eta_{\sigma_{(110)}}=1$). For the A$_1$ and A$_2$ gap functions, the EAZ class for the node structure is AIII, and there is no point node. On the other hand, for the B$_1$ gap function, the EAZ class is D for $\eta_I =1$ and $C$ for $\eta_I =-1$. Therefore, we can avoid a point node when the parity of the gap function is odd.

\bibliography{multipole}

\end{document}